\documentclass[aps,pre,reprint,superscriptaddress,longbibliography]{revtex4-1}

\usepackage{comment}
\usepackage{graphicx}
\usepackage{latexsym}
\usepackage{xcolor}
\usepackage{amsmath}
\usepackage{multirow}
\usepackage{booktabs}
\usepackage{placeins} 
\usepackage{float}
\usepackage{ulem} 
\graphicspath{{images/}}

\renewcommand{\vec}[1]{\mathbf{#1}}

\newcommand{\kB}{k_\mathrm{B}}

\newcommand{\unit}[1]{\vec{\widehat{#1}}}
\newcommand{\dif}{\text{d}}

\newcommand{\be}{\begin{equation}}
\newcommand{\ee}{\end{equation}}
\newcommand{\bea}{\begin{eqnarray}}
\newcommand{\eea}{\end{eqnarray}}

\begin{document}

\title{Twist-bend coupling and the statistical mechanics \\ 
of the twistable worm-like chain model of DNA: \\
perturbation theory and beyond}

\author{Stefanos\ K.\ Nomidis}
\affiliation{KU Leuven, Institute for Theoretical Physics, Celestijnenlaan
200D, 3001 Leuven, Belgium}
\affiliation{Flemish Institute for Technological Research (VITO), Boeretang 200, 
B-2400 Mol, Belgium}

\author{Enrico\ Skoruppa}
\affiliation{KU Leuven, Institute for Theoretical Physics, Celestijnenlaan
200D, 3001 Leuven, Belgium}

\author{Enrico\ Carlon}
\affiliation{KU Leuven, Institute for Theoretical Physics, Celestijnenlaan
200D, 3001 Leuven, Belgium}

\author{John\ F.\ Marko}
\affiliation{Department of Physics and Astronomy, and Department of
Molecular Biosciences, Northwestern University, Evanston, Illinois 60208,
USA}

\date{\today}

\begin{abstract}
The simplest model of DNA mechanics describes the double helix as a continuous 
rod with twist and bend elasticity. Recent work has discussed the relevance of a 
little-studied coupling $G$ between twisting and bending, known to arise from 
the groove asymmetry of the DNA double helix. Here, the effect of $G$ on the 
statistical mechanics of long DNA molecules subject to applied forces and 
torques is investigated. We present a perturbative calculation of the effective 
torsional stiffness $C_\text{eff}$ for small twist-bend coupling. We find that 
the ``bare'' $G$ is ``screened'' by thermal fluctuations, in the sense that the 
low-force, long-molecule effective free energy is that of a model with $G=0$, 
but with long-wavelength bending and twisting rigidities that are shifted by 
$G$-dependent amounts. Using results for torsional and bending rigidities for 
freely-fluctuating DNA, we show how our perturbative results can be extended 
to a non-perturbative regime. These results are in excellent agreement with 
numerical calculations for Monte Carlo ``triad'' and molecular dynamics 
``oxDNA'' models, characterized by different degrees of coarse-graining, 
validating the perturbative and non-perturbative analyses. While our theory is 
in generally-good quantitative agreement with experiment, the predicted 
torsional stiffness does systematically deviate from experimental data, 
suggesting that there are as-yet-uncharacterized aspects of DNA 
twisting-stretching mechanics relevant to low-force, long-molecule mechanical 
response, which are not captured by widely-used coarse-grained models.
\end{abstract}


\maketitle

\section{Introduction}

In vivo, double-stranded DNA is typically found in a highly-deformed
state, which is in part due to the interaction with the many proteins
that bend and twist the double helix, but in part due to thermally-driven
deformations.  A substantial effort has been devoted to the study of
many aspects of DNA mechanics, such as its response to applied twist
and bending deformations \cite{mark15}. These studies often rely on
homogeneous elastic models, which, despite their simplicity, describe
many aspects of single-molecule experiments \cite{bust94, stri96,
moro97, bouc98, mosc09}, and are widely used to describe mechanical and
statistical-mechanical properties of DNA (see e.g.\ Refs.~\cite{guer17,
bard18, whit18, lian18, bleh18} ).

One of the simplest models describing DNA deformations is the twistable 
wormlike chain (TWLC), which describes the double helix as an inextensible
rod, for which twist and bend deformations are independent. Symmetry
arguments suggest that the TWLC is incomplete: The inherent asymmetry
of the DNA molecular structure, with its major and minor grooves, gives
rise to a coupling $G$ between twisting and bending~\cite{mark94}. Only
a limited number of studies have considered the effect of twist-bend
coupling on DNA mechanics~\cite{moha05, nomi17, skor17, skor18}. A
systematic analysis of coarse-grained models with and without groove
asymmetry has highlighted several effects associated with twist-bend
coupling at long~\cite{skor17} and short~\cite{skor18} length scales.
Here, we aim to clarify the role of $G$ in the statistical mechanics of
long DNA molecules, as analyzed in optical and magnetic tweezers.

We focus on analytical and numerical results for the stretching and
torsional response of DNA with twist-bend coupling interaction $G\neq
0$. We first present a perturbative expansion for the partition function
of the molecule, in which $G$ is treated as the small parameter. The
lowest-order results show that twist-bend coupling softens the torsional
and bending stiffnesses of the double helix, recovering prior results
from entirely different calculations~\cite{nomi17}. Our new calculations
reveal the existence of a previously-unidentified large force scale
$f_0$; for forces below this scale, the bare elastic constants -
including $G$ - are not directly accessible in stretching and twisting
experiments. Instead, for forces below $f_0$, only renormalized bending
and twisting stiffnesses - which do depend on $G$ - are observed.
Because $f_0 \approx 600$~pN, the renormalized elastic model - which
is the $G=0$ TWLC - will be observed in essentially all conceivable
single-molecule experiments. Thus, $G$ is ``screened'', effectively
renormalized to $G=0$, in single-molecule DNA mechanics experiments.

Prior work~\cite{nomi17} suggests a strategy to generalize our results
beyond perturbation theory, to the regime where DNA is stretched by forces
less than $f_0$.  We validate both the perturbative and non-perturbative
results using numerical calculations corresponding to commonly-used
coarse-grained DNA elasticity models; our results turn out to closely
describe results of those numerical models.  Given this validation,
we turn to experimental data which are reasonably well described by the
low-force model, but for which there remain discrepancies, suggesting
effects beyond simple harmonic elastic models like the TWLC.

\section{Elasticity models of DNA}

\begin{figure}[t]
\centering\includegraphics[width=0.7\linewidth]{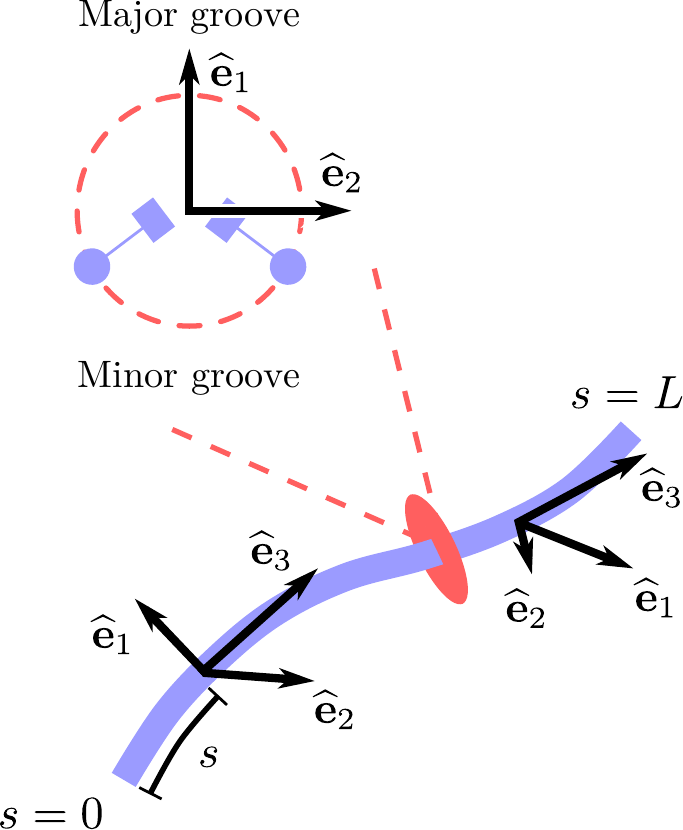}
\caption{Bottom: The configuration of a twistable elastic rod can be
mathematically described with an orthonormal set of vectors $\{ \unit e_1,
\unit e_2, \unit e_3 \}$ assigned to every point~$s$ along the rod. The
vector $\unit e_3$ is the tangent to the curve, and describes the bending
fluctuations along the rod. Top: Cross-section of the rod, indicating
how the remaining vectors $\unit e_1$ and $\unit e_2$, which describe
the torsional state, may be chosen in the particular case of DNA.}
\label{Fig:twlc}
\end{figure}

To describe the conformation of a continuous, inextensible, twistable
elastic rod, one can associate a local orthonormal frame of three unit
vectors $\{\unit e_i\}$ ($i=1,2,3$) with every point along the rod
(Fig.~\ref{Fig:twlc}). In a continuous representation of DNA, the common
convention is to choose $\unit e_3$ tangent to the curve and $\unit e_1$
pointing to the DNA major groove. The frame is completed with a third vector,
defined as $\unit e_2 = \unit e_3 \times \unit e_1$. An unstressed B-form
DNA corresponds to a straight, twisted rod, with the tangent $\unit e_3$
being constant, and with $\unit e_1$ and $\unit e_2$ rotating uniformly
about it, with a full helical turn every $l\approx3.6$~nm, or equivalently
every $10.5$ base pairs.

Any deformation from this unstressed configuration can be described by a 
continuous set of rotation vectors $\vec \Omega$ connecting adjacent local 
frames $\{\unit e_i\}$ along the rod, using the differential equation
\begin{equation} \label{app:diffeq2}
\frac{\dif \unit e_i}{\dif s} =  
\left(\vec{\Omega} + \omega_0 \unit e_3 \right) \times  \unit e_i,
\end{equation}
where the internal parameter $s$ denotes the arc-length coordinate
(Fig.~\ref{Fig:twlc}), and $\omega_0=2\pi/l \approx 1.75$~nm$^{-1}$ is
the intrinsic twist of DNA. Upon setting $\vec\Omega=\vec0$, one obtains
the unstressed configuration mentioned above.  Thus, a nonzero rotation
vector $\vec\Omega(s) \neq \vec0$ corresponds to a local deformation
at $s$ around this ground state.  Defining $\Omega_i \equiv \unit e_i
\cdot \vec\Omega$, it follows that $\Omega_1(s)$ and $\Omega_2(s)$
describe local bending deformations, while $\Omega_3(s)$ describes
twist deformations. In the remainder of the paper the $s$-dependence of
$\vec\Omega$ will be implicit.

Symmetry analysis of the DNA molecule requires the energy functional
$E$ to be invariant under the transformation $\Omega_1 \to - \Omega_1$,
with the consequence that~\cite{mark94}
\begin{equation}
\beta E = \frac{1}{2} \int_0^L \dif s \left( 
A_1 \Omega_1^2 + A_2 \Omega_2^2 + C \Omega_3^2 + 2G \Omega_2 \Omega_3 
\right), 
\label{mod_any}
\end{equation} 
where $\beta \equiv 1/\kB T$ is the inverse temperature, $A_1$ and
$A_2$ the bending stiffnesses, $C$ the torsional stiffness and $G$
the twist-bend coupling constant. These coefficients have dimensions
of length, and can be interpreted as the contour distance along the
double helix over which significant bending and twisting distortions
can occur by thermal fluctuations. Our perturbative calculation will
use the isotropic-bending version of this model ($A_1=A_2=A$), which is
described by the following energy functional
\begin{equation}
\beta E = \frac{1}{2} \int_0^L \dif s \left[ 
A  \left(\frac{\dif\unit{e}_3}{\dif s} \right)^2  
+ C \Omega_3^2 + 2G \Omega_2 \Omega_3 
\right].
\label{mod} 
\end{equation} 
Here, we have used Eq.~\eqref{app:diffeq2} to express the sum $\Omega_1^2
+ \Omega_2^2$ as the derivative of the tangent vector. The TWLC is
obtained by setting $G=0$ in Eqs.~\eqref{mod_any} and \eqref{mod},
corresponding to the anisotropic and isotropic cases, respectively.

\section{Effective torsional stiffness}

\begin{figure}[t]
\centering\includegraphics[width=0.6\linewidth]{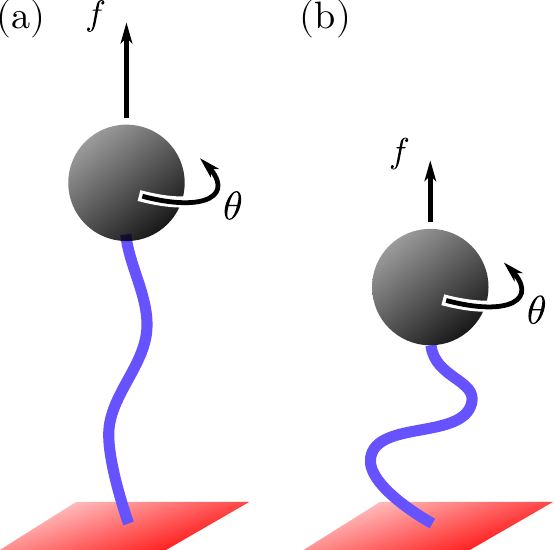}
\caption{Typical setup of a magnetic tweezers experiment. A DNA molecule
is covalently bound to a substrate at one end and to a paramagnetic bead
at the other end. An applied force $f$ stretches the molecule, while an
applied rotation $\theta$ twists it. The effective torsional stiffness
$C_\text{eff}$ is the proportionality constant connecting the applied
rotation with the exerted torque [Eq.~\eqref{eq:mtt}]. 
$C_\text{eff}$ increases with the force, hence it is
higher in (a) than in (b).
}
\label{Fig:MT}
\end{figure}

In a typical magnetic tweezers experiment, a single DNA molecule of $10^3-10^4$ 
bases is attached to a solid substrate and to a paramagnetic bead at its two 
ends (Fig.~\ref{Fig:MT}). The molecule can be stretched by a linear force $f$ 
and over- or undertwisted by an angle $\theta$.  The resulting torque~$\tau$ 
exerted by the bead, which can be experimentally measured~\cite{lipf10, kaue11, 
ober12, lipf14}, is linear in $\theta$ for small $\theta$
\begin{equation}\label{eq:mtt}
\tau \approx \frac{\kB T C_\text{eff}}{L} \theta.
\end{equation}
Here $C_\text{eff}$ is the \emph{effective} torsional stiffness (in contrast to 
the intrinsic stiffness C), and represents the central quantity of interest 
here.  It expresses the resistance of the DNA to a global torsional deformation, 
applied at its two ends.

As discussed in more detail below, $C_\text{eff}$ is in general lower than its 
intrinsic equivalent $C$. More specifically, at low stretching forces the 
bending fluctuations can absorb a significant part of the applied torsional 
stress, leading to a globally-reduced torsional resistance $C_\text{eff} < C$.  
On the other hand, when the applied force is sufficiently large, bending 
fluctuations are mostly suppressed, and hence the effective torsional stiffness 
tends to approach the intrinsic one.  As a consequence, $C_\text{eff}$ is going 
to be a monotonically-increasing function of the stretching force.

Moroz and Nelson derived an expression of $C_\text{eff}$ for the TWLC in the 
limit of high forces~\cite{moro97, moro98}. In spite of the good qualitative 
agreement between the theory and early experiments, more recent studies reported 
systematic deviations~\cite{lipf10, lipf11, lipf14, nomi17}. For completeness we 
will first present in Sec.~\ref{sec:Ceff_TWLC} a short derivation of the 
TWLC-based theory by Moroz and Nelson. The pertubative calculation in small $G$ 
is discussed in Sec.~\ref{sec:Ceff_MS} and generalized beyond perturbative 
expansion in Sec.~\ref{sec:Ceff_nonpert}.

\subsection{The TWLC limit ($G=0$)}
\label{sec:Ceff_TWLC}

Moroz and Nelson~\cite{moro97, moro98} mapped the twisted and stretched TWLC 
onto a quantum mechanical problem of a spinning top, and $C_\text{eff}$ was 
obtained from the ground state of the associated Schr\"odinger equation. Here we 
present an alternative derivation, following the scheme illustrated in 
Ref.~\cite{mark98,mark15}, which proves to be more convenient for the 
perturbative calculation in small $G$. The starting point is the partition 
function of a TWLC under applied force $f$ and torque $\tau$. The latter induces 
a rotation by an angle $\theta$ on the end point of the molecule 
(Fig.~\ref{Fig:MT}). The excess linking number, which we will use throughout 
this work, is $\Delta\text{Lk} = \theta/2\pi$. 

To calculate the partition function, we integrate over all possible 
configurations of the twistable rod, which can be parametrized by the tangent 
vector $\unit e_3(s)$ and the twist density $\Omega_3(s)$. The resulting path 
integral takes the form:
\begin{equation}
Z_0 = \int{\cal D} [\unit e_3,\Omega_3] \,
\, e^{\displaystyle -\beta E_0 + \beta \vec{f} \cdot \vec{R} 
+ 2\pi\beta \tau \Delta \text{Lk}},
\label{eq:Z0}
\end{equation}
where $E_0$ is the energy of the TWLC, obtained from Eq.~\eqref{mod} by setting 
$G=0$, and $\vec R$ the end-to-end vector
\begin{equation}
\vec R = \int_0^L \unit e_3 \, \dif s.
\label{def:endend}
\end{equation}
We assume that the force is oriented along the z-direction, hence $\vec f = f 
\unit z$, with $\unit z$ a unit vector. Using a result due to 
Fuller~\cite{full78}, originally derived for closed curves, 
the linking number can be expressed as the 
sum of twist and writhe, i.e.\ $\Delta \text{Lk} = \text{Tw} + \text{Wr}$. The 
excess twist is obtained by integrating over the twist density
\begin{equation}
\text{Tw} = \frac{1}{2\pi} \int_0^L \Omega_3(s) \, \dif s, 
\label{def:twist}
\end{equation}
while the writhe is given by~\cite{mark15}
\begin{equation}
\text{Wr} = \frac{1}{2\pi} \int_0^L \frac{\unit z \cdot (\unit e_3 \times 
\dif \unit e_3/\dif s)}
{1+\unit e_3\cdot\unit z} \ \dif s.
\label{def:writhe}
\end{equation}
This representation of the writhe as a single integral 
(and not as a double, nonlocal integral)
is correct modulo 1, with the integer portion equal to zero if
the molecule is sufficiently stretched so that it is unlikely to
loop back opposite to the direction of the applied force (the case of
interest here). 
Under these conditions, the denominator 
$1+\unit e_3\cdot\unit z$ does not vanish, and the integral in 
Eq.~\eqref{def:writhe} yields a finite value. 
The applicability of either Eq.~\eqref{def:writhe}
[or the double-integral version of it, which does not require the mod 1 of
Eq.~\eqref{def:writhe}] for a highly-stretched open chain has been
discussed and justified for extended polymers in prior works
\cite{mark95,volo97,moro97,bouc98,moro98}.

Next, we insert Eqs.~\eqref{def:endend},~\eqref{def:twist} and~\eqref{def:writhe} into 
Eq.~\eqref{eq:Z0}, and consider the limit of strong forces and weak torques. The 
partition function [Eq.~\eqref{eq:Z0}] reduces to a Gaussian in this limit, and 
can be easily estimated (details can be found in Appendix~\ref{app:twlc}). To 
lowest order in $\tau$ and at large forces, one obtains the following free 
energy
\begin{equation}\label{eq:Ftwlc}
F_0(f,\tau) = - f L + \sqrt{\frac{f \kB T}{A}} L 
              - \frac{\beta \tau^2 L}{2 C_\text{eff}} + \ldots,
\end{equation}
where the dots denote constant or higher-order terms in $\tau$. The effective 
torsional stiffness~$C_\text{eff}$ is given by (see Appendix~\ref{app:twlc})
\begin{equation}
\frac{1}{C_\text{eff}} =  \frac{1}{C} +
\frac{1}{4A}\sqrt{\frac{\kB T}{fA}} + \ldots .
\label{eq:ceff}
\end{equation}
This equation was originally derived by Moroz and Nelson~\cite{moro97,
moro98} in the fixed-torque ensemble. The same
expression can also be obtained in the fixed-linking-number ensemble
\cite{bouc98,jeon08}. At high forces, Eq.~\eqref{eq:ceff} approaches
the twisted-rod limit and $C_\text{eff} \to C$, but, in general, bending
fluctuations soften the DNA torsional stiffness, so that $C_\text{eff}
< C$. The latter originates from a global coupling between torque and
writhe (not to be confused with the local twist-bend coupling considered
below). Note that the effect of bending fluctuations is governed by
the dimensionless parameter $\sqrt{\kB T/fA}$, which is small at low
temperatures or high forces.

\subsection{Perturbative (small-$G$) expansion}
\label{sec:Ceff_MS}

We now construct a perturbation expansion for the partition function, using  $G$ 
as the small parameter (the length scale determining whether $G$
is ``small'' will be made clear below). The full partition function is
\begin{equation}
Z = \int{\cal D} [\unit e_3,\Omega_3] \,
\, e^{\displaystyle -\beta E + \beta \vec{f} \cdot \vec{R} + 
\beta \tau 2\pi \Delta \text{Lk}},
\label{eq:Z}
\end{equation}
where now $\beta E$ is given by Eq.~\eqref{mod} and contains a twist-bend 
coupling term. Assuming that $G$ is small, we can expand the Boltzmann factor in 
powers of $G$, which gives to lowest order:
\begin{eqnarray}
Z &\approx& Z_0 \left[ 1 + \frac{G^2}{2}  
\left\langle \left( \int_0^L \Omega_2 \Omega_3 \ \dif s
\right)^2\right\rangle_0 + \ldots \right],
\label{eq:pert}
\end{eqnarray}
where $\langle . \rangle_0$ denotes the average with respect to the unperturbed 
(TWLC) partition function [Eq.~\eqref{eq:Z0}]. Note that in the perturbative 
expansion, the term linear in $G$ vanishes by the $\Omega_2 \to - \Omega_2$ 
symmetry of the TWLC. The full calculation of the average in the right-hand side 
of Eq.~\eqref{eq:pert} is given in Appendix~\ref{appC}. The final expression for 
the free energy is of the form [Eq.~\eqref{app:Fpert}]
\begin{equation}
F(f,\tau) = -f L + \sqrt{\frac{f\kB T}{A^*}} L
+ \Gamma \tau L - \frac{\beta \tau^2 L}{2 C_\text{eff}}
+ \ldots,
\label{eq:free_en_pert}
\end{equation}
where terms of negligible contribution were omitted (see
Appendix~\ref{appC}).  

In the above, we have introduced the rescaled bending stiffness
\begin{equation}
\frac{1}{A^*} = \frac{1}{A} \left(1 + \frac{G^2}{2AC} \right),
\label{astar}
\end{equation}
together with the parameter
\begin{equation}
\Gamma = \frac{G^2 d^2(f)}{8A^2C^2 \omega_0}.
\label{defGamma}
\end{equation}
The dimensionless, force-dependent scale factor $d(f)$ will be discussed below;
we note that it appears in the coefficient $\Gamma$ [Eq.~\eqref{defGamma}], but
not in $A^*$ [Eq.~\eqref{astar}]. Finally, the form of the effective torsional 
stiffness $C_\text{eff}$ is
\begin{equation}
\frac{1}{C_\text{eff}} = \frac{1}{C} \left[ 1 + \frac{G^2}{AC}d(f) \right]
      + \frac{1}{4A} \left( 1 + \frac{3}{4} \frac{G^2}{AC} \right)
\sqrt{\frac{\kB T}{fA}}.
\label{eq:ceff_pert}
\end{equation}
The scale factor $d(f)$ is also present in $C_\text{eff}$.

Examination of these formulae indicate that the expansion is in powers of the 
dimensionless parameter $G^2/AC$, which, given our current estimates for the 
stiffnesses ($A \approx 50$ nm, $C \approx 100$ nm, $G \approx 30$ nm), is less 
than 1, although we note that $G^2/AC< 1$ is a stability requirement for the 
microscopic energy \cite{mark94}.  Our computation neglects terms beyond first 
order in $G^2/AC$.

\subsection{Effective torsional stiffness $C_\text{eff}$}

Equation~\eqref{eq:ceff_pert} is the central result of this paper, and extends 
the TWLC result by Moroz and Nelson [Eq.~\eqref{eq:ceff}], which is recovered in 
the limit $G \to 0$. The perturbative corrections are governed by the 
dimensionless parameter $G^2/AC$, and give rise to a further torsional softening 
of the molecule, i.e.\ $C_\text{eff} (G\neq0) < C_\text{eff} (G=0)$, as pointed 
out in Ref.~\cite{nomi17}. Equation~\eqref{eq:ceff_pert} contains also a 
force-dependent, crossover function, which can be approximated as (see 
Appendix~\ref{appC})
\begin{equation}
 d(f) \approx \frac{1}{1+f/f_0},
\end{equation}
where $f_0=A\omega_0^2 \kB T$ is the characteristic force above which $d(f)$ 
starts to significantly drop below its low-force limit of $d(0)=1$. To 
understand this force scale, which has no counterpart in the Moroz and Nelson 
formula [Eq.~\eqref{eq:ceff}], we recall that the correlation length for a 
stretched wormlike chain is $\xi = \sqrt{A\kB T/f}$~\cite{mark15}. Therefore, 
$f_0$ is the force associated with a correlation length of the order of the 
distance between neighboring bases, i.e.\ $\xi=1/\omega_0$. 

For DNA ($A\approx 50$~nm, $\omega_0 = 1.75$~nm$^{-1}$, $\kB T=4$~pN$\cdot$nm) 
we get $f_0 \approx 600$~pN, which is far above the force where the double 
helix starts to be itself stretched ($\approx 20$ pN), force-denatured 
($\approx 
60$~pN), and is in fact comparable to where the covalently-bonded backbones 
will break. Hence, for forces relevant to experiments we are concerned with ($f 
< 10$~pN), one may simply set $d(f)\approx 1$. We will refer to this limit as 
the ``low-force limit'', but one should keep in mind that our perturbative 
theory is computed for the ``well-stretched" limit, i.e.\ $f > \kB T/A \approx 
0.1$ pN. Therefore our perturbative theory is applicable in the force range of 
roughly $0.1$ to $10$~pN.

We emphasize that Eq.~\eqref{eq:ceff_pert} can be written as
\begin{equation}
\frac{1}{C_\text{eff}} = \frac{1}{C^*} + 
\frac{1}{4A^*} \sqrt{\frac{\kB T}{f A^*}},
\label{eq:rescal_Ceff}
\end{equation}
where $A^*$ is defined in~\eqref{astar}, and
\begin{equation}
\frac{1}{C^*} \equiv \frac{1}{C} \left[1 + \frac{G^2}{AC} \, d(f) \right]
+{\cal O}(G^2/AC).
\label{cstar}
\end{equation}
Equation~\eqref{eq:rescal_Ceff} has exactly the same form as the Moroz and 
Nelson formula [Eq.~\eqref{eq:ceff}], with rescaled bending and torsional 
stiffnesses. The importance of this result is paramount: since $d(f) = 1$ in the 
range of experimentally relevant forces, the torsional stiffness (and in fact 
the partition function itself) depends only on the ``renormalized'' stiffnesses 
$A^*$ and $C^*$, meaning that $G$ by itself cannot be determined from fitting of 
$C_\text{eff}(f)$ (or any other equilibrium quantity versus $f$); only the 
effective stiffnesses $A^*$ and $C^*$ can be determined from experiments in the 
low-force regime.

\subsection{Non-perturbative result for $C_{\rm eff}$ valid for $f < f_0$}
\label{sec:Ceff_nonpert}

Equation~\eqref{eq:ceff_pert} has been derived on the basis of a systematic 
perturbation expansion in $G$ (more formally, in the small parameter $G^2/AC < 
1$). When cast in the form of Eq.~\eqref{eq:rescal_Ceff}, it is apparent that 
there is a simple way to extend the results to a more general, nonperturbative 
case, where $G$ may be large and the bending possibly anisotropic, i.e.\ $A_1 
\neq A_2$ in Eq.~\eqref{mod_any}. The key physical idea here is that for forces 
below the gigantic force scale $f_0$, thermal fluctuations at the helix scale 
where $G$ correlates bending and twisting fluctuations are unperturbed ($d(f) = 
1$), and therefore we might as well just consider DNA to have the effective 
twisting and bending stiffnesses that it has at \emph{zero} force.

In absence of applied torques and forces ($f,\tau=0$), the partition
function of Eq.~\eqref{eq:Z} can be evaluated exactly~\cite{nomi17}. 
Due to twist-bend coupling, the bending and torsional stiffnesses are 
renormalized as~\cite{nomi17}:
\begin{eqnarray}
\kappa_\text b &=& A \, 
\frac
{\displaystyle  1-\frac{\varepsilon^2}{A^2}-\frac{G^2}{AC}
\left( 1 + \frac{\varepsilon}{A} \right)}
{\displaystyle 1-\frac{G^2}{2AC}},
\label{kappab} \\
\kappa_\text t &=& C \, 
\frac
{\displaystyle 1-\frac{\varepsilon}{A}-\frac{G^2}{AC}}
{\displaystyle 1-\frac{\varepsilon}{A}},
\label{kappat} 
\end{eqnarray}
where we have introduced the parameters $A=(A_1+A_2)/2$ and $\varepsilon = 
(A_1-A_2)/2$. {Eqs.~\eqref{kappab} and \eqref{kappat} quantify the energetic 
cost of bending and twisting deformations, respectively, in the same way $A$ and 
$C$ do within the TWLC.  Note that, by setting $G=0$ in these expressions, one 
recovers the TWLC limit, $\kappa_\text{t}=C$ and 
$\kappa_\text{b}=2A_1A_2/(A_1+A_2)$ i.e.\ the renormalized bending stiffness is 
the harmonic mean of $A_1$ and $A_2$, which is a known result (see e.g.\ 
Refs.~\cite{lank03,esla09}).  If $G \neq 0$  one has $\kappa_\text{b}< 
2A_1A_2/(A_1+A_2)$ and $\kappa_\text{t}<C$, i.e.\ twist-bend coupling 
``softens'' the bending and twist deformations of the DNA molecule, already if 
$f=0$, $\tau=0$.  Eq.~\eqref{eq:ceff_pert} then describes two different effects: 
one is the thermally-induced torsional softening due to bending fluctuations 
[already present in the TWLC expression~\eqref{eq:ceff}] and the other is the 
$G$-induced softening, which is captured by the two factors between parentheses 
in Eq.~\eqref{eq:ceff_pert}.

Setting $\varepsilon=0$ and expanding Eqs.~\eqref{kappab} and
\eqref{kappat}, one finds $\kappa_\text b = A^* + \mathcal O (G^4)$ and
$\kappa_\text t = C^* + \mathcal O(G^4)$, which suggests the following,
more general, nonperturbative result for $C_\text{eff}$, valid as long
as $f \ll f_0$
\begin{equation}
\frac{1}{C_\text{eff}} = \frac{1}{\kappa_\text t} + 
\frac{1}{4\kappa_\text b} \sqrt{\frac{\kB T}{f \kappa_\text b}}.
\label{eq:ceff_nonp}
\end{equation}
This relation, similar to Eq.~\eqref{eq:rescal_Ceff}, has the same form as the 
Moroz and Nelson formula [Eq.~\eqref{eq:ceff}], with $A$ and $C$ replaced by 
$\kappa_\text b$ and $\kappa_\text t$ [much as our result for $C_\text{eff}$ of 
Eq.~\eqref{cstar} has the Moroz-Nelson form with $A\rightarrow A^*$ and 
$C\rightarrow C^*$]. As we will show in the next Section, this new, 
nonperturbative result for the continuum model is in excellent agreement with 
numerical Monte Carlo (MC) and molecular dynamics (MD) calculations.

\subsection{Twist-bend-coupling-induced DNA unwinding}

An intriguing feature of the perturbative calculation is the appearance of a 
term linear in $\tau$ in the free energy [Eq.~\eqref{eq:free_en_pert}], which 
induces an unwinding of the helix at zero torque. In particular, from 
Eqs.~\eqref{eq:Z}, \eqref{eq:free_en_pert} and \eqref{defGamma} it follows that
\begin{equation}
\left. \langle \Delta \text{Lk} \rangle \right|_{\tau=0} = 
-\frac{1}{2\pi} \left. \frac{\partial F}
{\partial \tau}\right|_{\tau=0} 
= -\frac{d^2(f)}{16\pi\omega_0} \, 
\frac{G^2 L}{A^2 C^2}.
\label{temp_unwind}
\end{equation}
The scale for this thermal unwinding is very small: using typical values of DNA 
parameters ($A \approx 50$ nm, $C \approx 100$ nm, $G = 30$ nm, $\omega_0 = 
1.75$ nm$^{-1}$) we find an unwinding angle per contour length of $2\pi \left< 
\Delta {\rm Lk} \right>/L = -G^2/ 8 \omega_0 A^2 C^2 \approx -2.6\times 
10^{-6}$ rad/nm (about $-5 \times 10^{-5}$ degrees per base pair).  This 
$G$-generated shift in helix twisting is inconsequentially small, but it is 
worth noting that this term is present in the perturbation theory.

It has been long known that there is a gradual unwinding of the double helix as 
the temperature is increased \cite{depe75,dugu93}, and this effect has been 
recently observed at the single-DNA level \cite{krie18}. Although one might 
imagine an overall $T^2$ dependence of this term (from the factors of $\beta$ in 
the Boltzmann factor), this dependence can only generate a tiny fraction of the 
observed temperature-dependent unwinding of $\approx -1\times 10^{-2}$ 
degrees/K$\cdot$bp. The experimentally observed unwinding is likely due to 
temperature-dependence DNA conformational changes \cite{krie18}, and is beyond 
the scope of being captured by the simple elastic models discussed here; in 
particular the observed unwinding of DNA with increasing temperature is not 
attributable to the twist-bend coupling $G$.

\subsection{``Janus strip'' limit ( $\omega_0 \to 0$ )}

Equation~\eqref{eq:ceff_nonp} is not generally valid for any arbitrary polymer 
with twist-bend coupling, but its validity is linked to the physical parameters 
characterizing DNA elasticity.  These conspire to set forces encountered in 
typical experiments (below 10 pN) to be far below the characteristic force $f_0 
= \kB T A \omega_0^2 \approx 600$ pN at which one starts to see effects at the 
helix repeat scale, i.e.\ force-driven unwinding of the double helix due to 
quenching of thermal fluctuations and the influence of $G$. In this sense, 
$\omega_0$ can be regarded as a ``large'' parameter: combinations of it and the 
elastic constants give dimensionless constants large compared to unity, e.g.\ 
$A\omega_0$, $C \omega_0 \approx 10^2 \gg 1$  and $\sqrt{A\kB T/f} \omega_0 \gg 
1$ for $f < 10$~pN). 

For this reason, several $\omega_0$-dependent terms, which in principle would 
contribute to $C_\text{eff}$ at order $G^2$, can in practice be neglected in 
the application of the theory to DNA [see e.g.\ Eqs.~\eqref{c29} and 
\eqref{c30} in Appendix]. The neglect of these terms leads to $C_\text{eff}$ 
taking the simple form given by Eq.~\eqref{eq:rescal_Ceff}, in which $A^*$ and 
$C^*$ are the renormalized stiffnesses. 

While not relevant to DNA, we might imagine other polymer structures for which 
$\omega_0$ is not so large, i.e.\ where $\omega_0$ is closer in size to $1/A$ or 
$1/C$.  In this case one cannot ignore these additional terms, and $d(f) \approx 
1/(1+f/f_0)$ might drop significantly over experimentally-relevant force 
ranges. Chiral proteins, lipid filaments, or even nanofabricated objects might 
comprise realizations of such situations.   

As an example we consider the extreme limit $\omega_0 \to 0$, corresponding to 
a ``Janus strip'', an elastic strip with inequivalent faces (i.e.\ 
inequivalent major and minor ``grooves''), and, thus, nonzero $G$. In this 
case, using the more complete and complicated results for the perturbative 
expansion given in the Appendix, we obtain to lowest order in $G$
\begin{equation}
\frac{1}{C_\text{eff}} = 
\frac{1}{C} + \frac{1}{4A} 
\left[ 1 + \left(\frac{3}{4} + \frac{2A^2}{C^2} \right) 
\frac{G^2}{AC} 
\right]
\sqrt\frac{\kB T}{fA}.
\label{eq:omega0_0}
\end{equation}
In this limit, compared to the large-$\omega_0$ case relevant to DNA,
there is a more gradual shift of $C_\text{eff}$ up to its high-force
limit, and an inequivalence of the form of $C_\text{eff}$ to the
Moroz-Nelson form. Physically, this is because the intrinsic chirality of
the filament is now gone, eliminating the ``screening'' of effects of $G$
at low forces, and the simple dependence of the low-force themodynamics
on only the coarse-grained stiffnesses $\kappa_\text t$ and $\kappa_\text
b$. Experiments on such Janus strips, or on ``soft-helix'' objects where
$A \omega_0$, $C \omega_0 < 1$ could provide realizations of this limit
of the theory.  Ref. \cite{nomi17} showed that Eq.~\eqref{eq:omega0_0}
fits experimental data for DNA surprisingly well, despite not taking
account of double helix chirality.

\section{Numerical Calculations}

To check the validity of the analytical results presented above, we performed 
numerical simulations of two different models.  The first model, referred to as 
the \emph{triad model}, is obtained from the discretization of the continuum 
elastic energy \eqref{mod_any} and treated using MC computations. The second 
model is oxDNA, a coarse-grained model of nucleic acids~\cite{ould10}, treated 
using MD calculations.

\subsection{Triad model}
\label{sec:triad}

The triad model is comprised of a series of $N$ orthonormal
vectors $\{\unit e_i (k)\}$ with $i=1,2,3$ and $k=0,1,2 \ldots N$,
each representing a single base pair, interacting with its neighbors
according to Eq.~\eqref{mod_any}. The total length of the molecule is
$L=Na$, with $a=0.34$~nm the base pair distance.  The ground state of
this model is a twisted, straight rod, with $\unit e_3$ being aligned
with the direction of the stretching force, and the vectors $\unit
e_1$, $\unit e_2$ rotating about $\unit e_3$ with an angular frequency
$\omega_0$. A cluster move consisted of a rotation of the whole subsystem
beyond a randomly-selected triad by a random angle. The new rotation
vector $\vec \Omega$ was calculated based on Rodrigues' rotation formula
(see e.g.\ Supplementary Material of Ref.~\cite{skor17}), then the energy
was updated from a discretized version of Eq.~\eqref{mod_any}, with the
addition of a force term [see Eq.~\eqref{eq:Z0}]. The move was accepted or
rejected according to the Metropolis algorithm. The stiffness constants
$A_1$, $A_2$, $C$ and $G$ are input parameters for the model, and may,
therefore, be arbitrarily chosen, provided the stability condition $G^2
< A_2 C$ is met [for which the quadratic form in Eq.~\eqref{mod_any}
is positive definite]

The effective torsional stiffness was calculated at zero torque from
linking number fluctuations:
\begin{equation}
 C_\text{eff} = \frac{L}{4\pi^2 \langle(\Delta \text{Lk} - 
\langle\Delta \text{Lk}\rangle)^2\rangle},
\label{measure_ceff}
\end{equation}
The variance of linking number in the denominator was evaluated
from the topological relation $\Delta \text{Lk} = \Delta\text{Tw} +
\text{Wr}$, with twist and writhe obtained from the discretization of
Eqs.~\eqref{def:twist} and \eqref{def:writhe}, respectively. To check
the validity of our results, the writhe was also evaluated from the
double-integral formula, following the method of Ref.~\cite{chou14},
and no significant differences were found for forces $> 0.25$~pN. In
all simulations, the size of the system was 600 triads (base pairs),
above which the results remained identical within that force range.

\begin{figure}[t!]
\centering\includegraphics{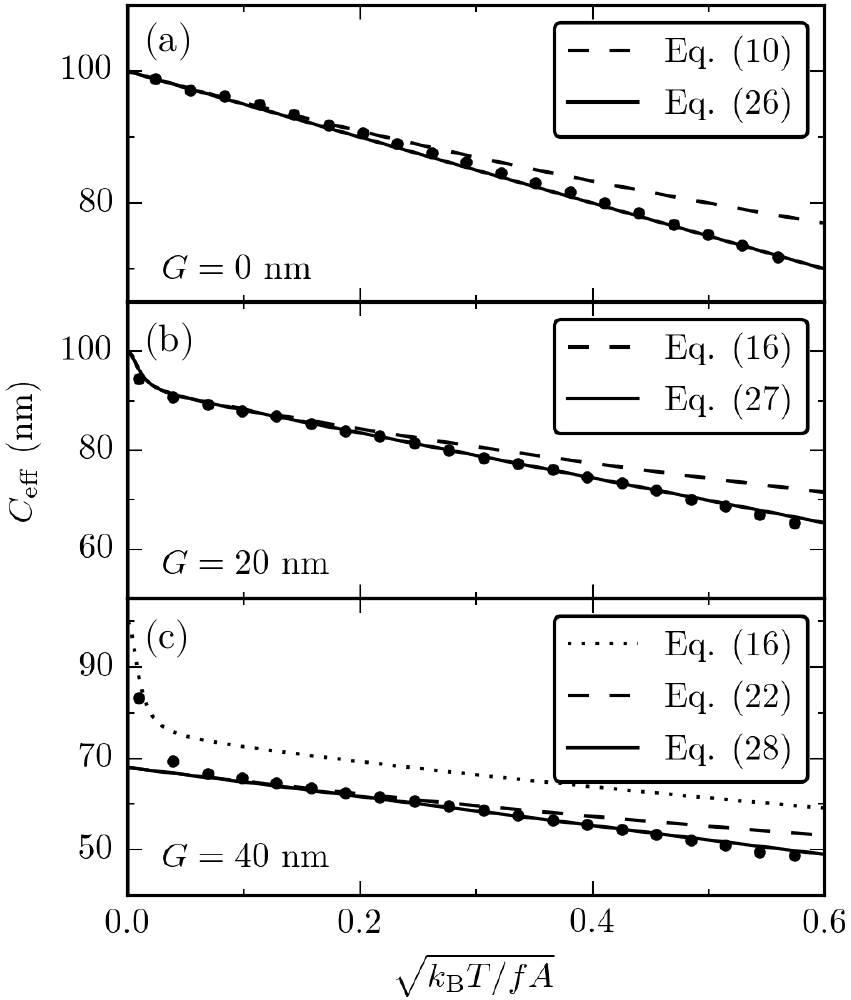}
\caption{Comparison of Monte Carlo simulations of the triad model
(points) with various analytical expressions of $C_\text{eff}$ (lines)
for $A=50$~nm, $C=100$~nm and (a) $G=0$, (b) $G=20$~nm and (c) $G=40$~nm. 
The data are plotted as
a function of $\sqrt{\kB T/fA}$ and correspond to $f \geq 0.25$~pN. The
numerical results are in excellent agreement with the analytical
expressions, both in the perturbative and nonperturbative regimes
(see text). Error bars of Monte Carlo data are smaller than symbol 
sizes.}
\label{Fig:Ceff}
\end{figure}

\subsubsection{Isotropic bending}

Figure~\ref{Fig:Ceff} shows the results of Monte Carlo calculations for
the isotropic model of Eq.~\eqref{mod}, with $A=50$~nm, $C=100$~nm and
$G=0, 20, 40$~nm (top to bottom). The data are plotted as a function of
the dimensionless parameter $\sqrt{\kB T/fA}$. The numerical errors are
smaller than the symbol sizes, and hence not shown.

In absence of twist-bend coupling ($G=0$, upper panel), the Monte
Carlo data are in excellent agreement with the Moroz-Nelson theory. We
compare them both to Eq.~\eqref{eq:ceff} (dashed line) and the following
expression (solid line)
\begin{equation}
C_\text{eff} = C \left( 1 - \frac{C}{4A} \sqrt{\frac{\kB T}{fA}} \right).
\label{eq:ceff_inv}
\end{equation}
The latter is obtained from the lowest-order expansion of
Eq.~\eqref{eq:ceff} in $\sqrt{\kB T/fA}$, and is a straight
line when plotted as a function of the rescaled variable of
Fig.~\ref{Fig:Ceff}. Eqs.~\eqref{eq:ceff} and \eqref{eq:ceff_inv}
coincide to leading order in $1/\sqrt{f}$, and any differences in the
two expressions only occur at low force scales, where higher-order
corrections become relevant.  Eq.~\eqref{eq:ceff_inv} fits the Monte
Carlo data over the whole range of forces analyzed ($f
\geq 0.25$~pN), while Eq.~\eqref{eq:ceff} deviates at low forces 
\footnote{redAlthough our analysis was restricted to the
lowest-order term in a high-force expansion, the excellent agreement
of the numerical data with Eq.~\eqref{eq:ceff_inv} gives some indication
of the nature of the next-order term. Starting from
\begin{equation}
\frac{1}{C_\text{eff}} = \frac{1}{C} + \frac{a_1}{\sqrt{f}} + 
\frac{a_2}{f},
\end{equation} 
one gets
\begin{equation} 
\frac{C_\text{eff}}{C} = 1 - \frac{a_1}{\sqrt{f}} + 
\left( a_1^2 C - a_2\right) \frac{C}{f}.
\end{equation} 
Numerics suggests that the term in $1/f$ is very small, implying
$a_2 \approx a_1^2 C$.}

The middle panel of Fig.~\ref{Fig:Ceff} shows Monte Carlo results
for $G=20$~nm (points), which we compare both to the results for the
perturbative expansion for $1/C_{\rm eff}$, Eq.~\eqref{eq:rescal_Ceff}
(dashed line) and the following similar expansion result for $C_{\rm eff}$
(solid line)
\begin{equation}
C_\text{eff} = \frac{C}{1+\dfrac{G^2}{AC} d(f)} 
\left( 1 - 
\frac{C}{4A} \frac{1+\dfrac{3G^2}{4AC}}{1+\dfrac{G^2}{AC} d(f)}
\sqrt{\dfrac{\kB T}{fA}} \right),
\label{eq:ceff_nonp_df}
\end{equation}
obtained by expanding Eq.~\eqref{eq:ceff_pert} to lowest order in
$1/f$. The latter is in excellent agreement with Monte Carlo data in the
whole range of forces considered, indicating that $G=20$~nm falls within
the range of validity of the perturbative calculation ($G^2/AC = 0.08$
in this case).  We note that the two perturbative expansion results
converge together at high forces ($[k_B T/ (Af)]^{1/2} \rightarrow 0$
and also show the upturn at the very highest forces associated with the
force-dependence of $d(f)$.

Finally, the lower panel of Fig.~\ref{Fig:Ceff} shows the results of
Monte Carlo simulations for $G=40$~nm.  The numerical data deviate
substantially from Eq.~\eqref{eq:ceff_pert} (dotted line), indicating
that $G$ is leaving the range of validity of the perturbative calculation
($G^2/AC=0.32$).  The remaining curves show the nonperturbative result
for $1/C_{\rm eff}$ Eq.~\eqref{eq:ceff_nonp} (dashed line), together with
the following nonperturbative expression for $C_\text{eff}$ (solid line)
\begin{equation}
C_\text{eff} = \kappa_\text t \left( 1 - 
\frac{\kappa_\text t}{4\kappa_\text b} 
\sqrt{\frac{\kB T}{f\kappa_\text b}} \right),
\label{eq:ceff_nonp_inv}
\end{equation}
the latter being in excellent agreement with Monte Carlo data for $f <
f_0$, validating the nonperturbative result (note that $[k_B T /(f_0
A)]^{1/2} \approx 0.01$).  For $f > f_0$, the upturn of $C_{\rm eff}$
towards the bare value of $C$ is apparent; this effect, while not given
by the nonperturbative results, is present in the perturbation expansion
results.  We conclude that our non-perturbative result indeed provides
a quantitative account of $C_{\rm eff}$ for $f < f_0$, where we expect
it to be valid.

\begin{figure}[t]
\centering\includegraphics{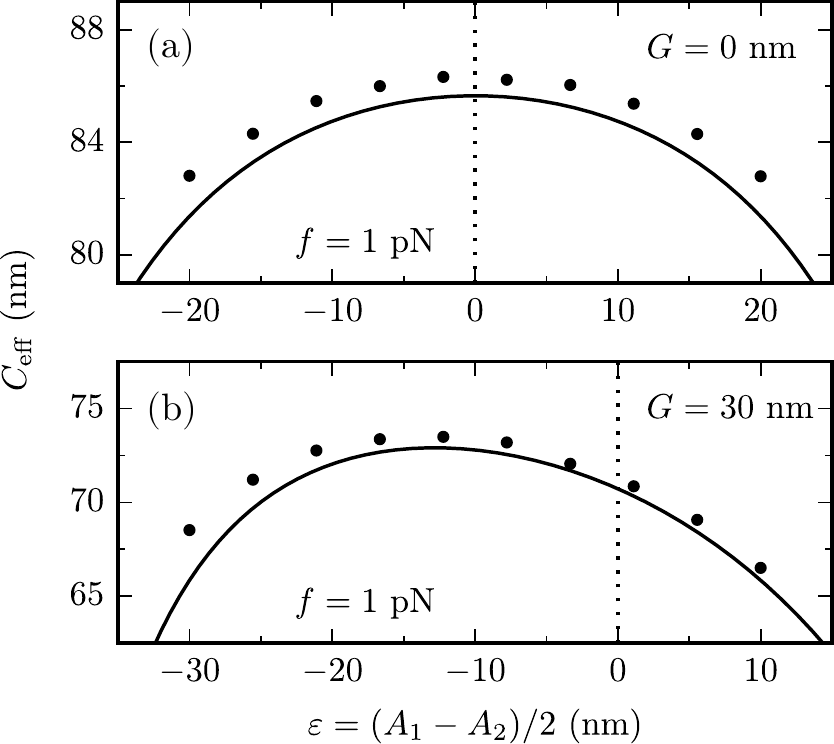}
\caption{Effect of bending anisotropy ($A_1 \neq A_2$) on the effective
torsional stiffness at a fixed force $f=1$~pN with (a) $G=0$ and (b)
$G=30$~nm. Numerical data from Monte Carlo simulations of the triad model
(points) are in good agreement with the analytical, nonperturbative
predictions of Eq.~\eqref{eq:ceff_nonp_inv} (solid lines). The vertical
dashed lines indicate the isotropic case ($\varepsilon = 0)$. A
nonvanishing twist-bend coupling (lower panel) induces a $\varepsilon
\to -\varepsilon$ symmetry breaking. Error bars of Monte Carlo data are
smaller than symbol sizes.}
\label{Fig:Ceff_f1}
\end{figure}

\subsubsection{Anisotropic bending}

While the perturbative calculation [Eq.~\eqref{eq:ceff_pert}] was
restricted to the isotropic case ($A_1=A_2$), the nonperturbative result
[Eq.~\eqref{eq:ceff_nonp_inv}] has a broader range of applicability, and
is able to describe the anisotropic case as well. Figure~\ref{Fig:Ceff_f1}
shows the results of Monte Carlo calculations of $C_\text{eff}$ for
various values of the anisotropy parameter $\varepsilon = (A_1-A_2)/2$
and a fixed value of the force $f=1$~pN. The Monte Carlo data are in
very good agreement with Eq.~\eqref{eq:ceff_nonp_inv}, plotted with
solid lines. The differences are within 5\%, and are probably due to
higher order corrections in $1/f$ (recall that all analytical results are 
based on a large-force expansion). In absence of twist-bend coupling
($G=0$), Eq.~\eqref{eq:ceff_nonp_inv} is symmetric in $\varepsilon$, as
in this case Eqs.~\eqref{kappab} and \eqref{kappat} give $\kappa_\text
b=(A^2-\varepsilon^2)/A$ and $\kappa_\text t=C$, respectively. A nonzero
$G$ induces nonvanishing terms, which are linear in $\varepsilon$
both in $\kappa_\text{b}$ and $\kappa_\text t$ [Eqs.~\eqref{kappab}
and \eqref{kappat}], leading to a breaking of the $\varepsilon \to
-\varepsilon$ symmetry.

\setlength{\tabcolsep}{6.7pt} 
\begin{table}[t]
\centering
\begin{tabular}{c c c c c c c }
\hline
\hline
\addlinespace[2pt]
& $A_1$ & $A_2$ & C & G & $\kappa_\text{b}$ & $\kappa_\text{t}$\\
\addlinespace[2pt]
\hline
\addlinespace[2pt]
oxDNA1 & 84(14) & 29(2) & 118(1) & $<$0.3 & 43 & 118 \\
oxDNA2 & 85(10) & 35(2) & 109(1) & 25(1)  & 44 &  92 \\
\addlinespace[2pt]
\hline
\hline
\end{tabular}
\caption{Values of the stiffness coefficients for oxDNA1 and
oxDNA2 (expressed in nm), derived from MD data of Ref.~\cite{skor17}.  
oxDNA1, which has symmetric grooves,
is characterized by a negligible twist-bend coupling constant,
while $G=25$~nm for oxDNA2, which has asymmetric grooves. 
The values of $\kappa_\text b$ and $\kappa_\text t$ are obtained from
Eqs.~\eqref{kappab} and \eqref{kappat}, respectively, and were found to
agree with direct computations of those quantities~\cite{skor17}.
Note that the oxDNA2 stiffness coefficients have been transformed to
coordinates compatible with this paper, see Appendix D).}
\label{table1}
\end{table}

\subsection{oxDNA}
\label{sec:oxdna}

oxDNA is a coarse-grained model describing DNA as two intertwined
strings of rigid nucleotides~\cite{ould10}. It has been used for the
study of a variety of DNA properties, ranging from single molecules to
large-scale complexes \cite{ould10, sulc12, snod15, skor17, skor18,
enge18}.  To date, two versions of oxDNA exist: one with symmetric
grooves (oxDNA1)~\cite{ould10} and one with asymmetric grooves
(oxDNA2)~\cite{snod15}. Comparing the torsional response of the two
versions will allow us to infer the effect of the groove asymmetry on
$C_\text{eff}$.  Differently from the triad model, in which the stiffness
constants $A_1$, $A_2$, $C$ and $G$ are input parameters, in oxDNA they
are determined by the molecular force fields used.  These force fields
were accurately tuned so that the experimental DNA structural, mechanical
and thermodynamic properties (as persistence length, melting temperatures
and torque-induced supercoiling) are well reproduced~\cite{ould10}.
As for real DNA, for oxDNA the elastic constants are emergent via
coarse-graining of fluctuations of smaller-scale, molecular motion
degrees of freedom.

The stiffness parameters of oxDNA were recently estimated from
the analysis of the equilibrium fluctuations of an unconstrained
molecule~\cite{skor17}, and are shown in Table~\ref{table1} (the
values of the elastic constants for oxDNA2 shown are the result of
transformation of the values obtained in Ref. \cite{skor17} for the
helical coordinate system used in that paper, to the non-helical
coordinate system of this paper; see Appendix~\ref{sec:int_bend}).
In line with the symmetry arguments of Ref.~\cite{mark94}, twist-bend
coupling is absent in oxDNA1 (symmetric grooves), while its magnitude
is comparable to that of the elastic constants $A_1$, $A_2$ and $C$ in
oxDNA2 (asymmetric grooves). Table~\ref{table1} also reports the values
of $\kappa_\text b$ and $\kappa_\text t$, which can be obtained in two
different, yet consistent, ways~\cite{skor17}: either indirectly from
Eqs.~\eqref{kappab} and \eqref{kappat}, by plugging in $A_1$, $A_2$,
$C$ and $G$ of Table~\ref{table1}, or directly from the analysis of the
corresponding correlation functions in simulations ($\kappa_\text b$
and $2\kappa_\text t$ are, respecively, the bending and twist stiffnesses
\cite{nomi17}).

\begin{figure}[t]
\centering\includegraphics{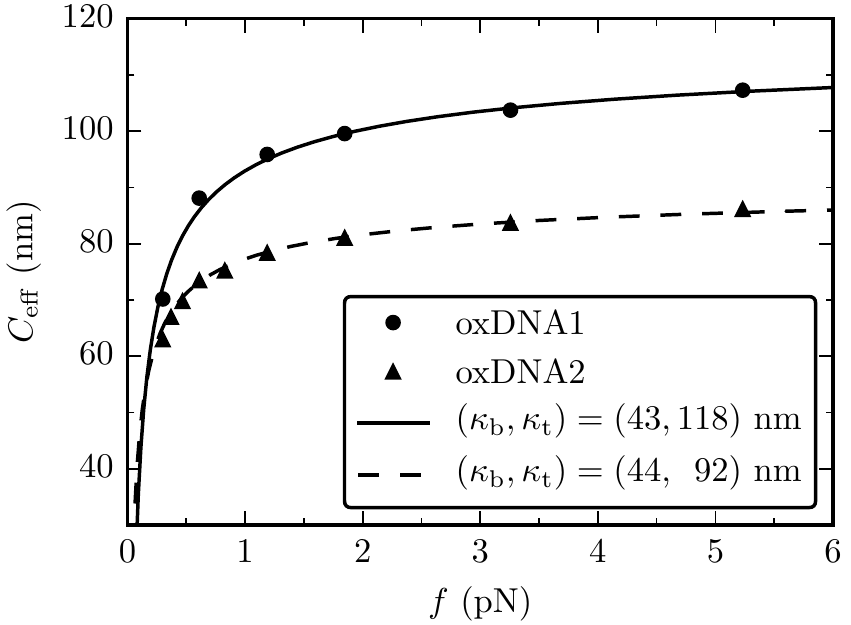}
\caption{Comparison of oxDNA simulations to Eq.~\eqref{eq:ceff_nonp_inv}.
Solid and dashed lines show the nonperturbative result for $C_{\rm eff}$
for oxDNA1 and oxDNA2 values of Table~\ref{table1}, respectively.
Error bars for the oxDNA data are smaller than symbol sizes.}
\label{Fig:oxdna} 
\end{figure}

Figure~\ref{Fig:oxdna} shows a plot of the effective torsional stiffness
as a function of the applied force, both for oxDNA1 (circles) and oxDNA2
(triangles). $C_\text{eff}$ was evaluated using twist fluctuations via
Eq.~\eqref{measure_ceff}. At large forces, and in agreement with the
experimental evidence, oxDNA undergoes a structural transition, hence
the simulations were restricted to $f \leq 10$~pN.  The solid and dotted
lines of Fig.~\ref{Fig:oxdna} are plots of Eq.~\eqref{eq:ceff_nonp_inv}
using $\kappa_\text b$ and $\kappa_\text t$ from Table~\ref{table1}.
For both oxDNA1 and oxDNA2 there is an excellent agreement between the 
nonperturbative
theory and simulations. In the case of oxDNA1 the nonperturbative theory reduces
to the Moroz-Nelson result, with $\kappa_{\rm t} = C$ and $\kappa_{\rm
b} = A(1-\epsilon^2/A^2)$; the good account of oxDNA1 $C_{\rm eff}$
by this formula was noted previously (see Ref.~\cite{mate15}, Fig. S7).
We note that in the light of our present results, this good agreement
validates the use of the values of the stiffness parameters obtained
in Ref.~\cite{skor17}.

\section{Discussion}

We have investigated the effect of the twist-bend coupling $G$ on the
statistical-mechanical properties of the twistable-wormlike-chain model
of a stretched DNA molecule, using
analytical and numerical methods.  Our major analytical results are
based on a  perturbative calculation of the effective torsional stiffness
$C_\text{eff}$, the torsional resistance of a long DNA molecule stretched
by an applied force $f$.  The calculation is valid for small values of
$G$, and generalizes the expression derived by Moroz and Nelson, which
was obtained for $G=0$ \cite{moro97}.

\subsection{Screening effect for $f < f_0$}

A striking feature of our theory is the appearance of a large force
scale, $f_0 = k_BT A \omega_0^2 \approx 600$~pN.  For forces well below
this gargantuan force level (essentially all single-DNA mechanical
experiments concern forces far below this value) the effect of $G$
becomes solely renormalization of the bending and twisting stiffnesses
$\kappa_{\rm b}$ and $\kappa_{\rm t}$; direct effects of twist-bend
coupling are ``screened" at lower force scales. Only at forces $f
\gg f_0$ do the bare elastic constants start to reveal themselves:
in this regime $C_\text{eff}$ finally approaches its intrinsic value
$C$. Note that the large force regime $f \gg f_0$
is experimentally inaccessible, as it corresponds to forces beyond
those where DNA rapidly breaks.

This ``screening" feature of the perturbative theory suggested to us
that we could consider DNA for $f < f_0$ to be described by a TWLC with
persistence lengths set to the zero-force long-molecule stiffnesses
$\kappa_{\rm b}$ and $\kappa_{\rm t}$.  Combining formulae for the
stiffnesses for freely fluctuating DNA~\cite{nomi17} with the $C_{\rm
eff}$ formula of Moroz and Nelson \cite{moro97} gave us a nonperturbative
formula for $C_{\rm eff}$ in terms of the elastic constants $A_1$, $A_2$,
$C$ and $G$.  MC calculations for the triad model which discretizes
the continuum elasticity theory \eqref{mod} were found to be in
excellent agreement both with the perturbative (i.e.,\ small $G$)
and nonperturbative (for larger $G$) expressions of $C_\text{eff}$.
We note that, despite being inaccessible experimentally, 
in MC simulations we observed the very high force behavior
of the perturbative theory - namely the increase of $C_{\rm eff}$ from
its low-force Moroz-Nelson behavior, towards its ``naked" value of $C$
in the triad MC calculations. 

The screening discussed here applies to single-molecule
measurements sampling the torsional response of a kilobase-long
molecule. Locally, at the distance of few base-pairs, twist-bend coupling
has directly-observable effects, as discussed recently \cite{skor18,cara18}.
For instance, in DNA minicircles, the twist oscillates as a response
to pure bending deformations, as seen in X-ray structures of nucleosomal
DNA \cite{skor18}.

\subsection{oxDNA under moderate forces is described by the TWLC plus
twist-bend coupling}

To test whether our analytical results describe the coarse-grained
behavior of a more realistic molecular model of DNA, we carried out
MD simulations of oxDNA, a coarse-grained model describing DNA as two
intertwined strings of rigid nucleotides~\cite{ould10}. $C_\text{eff}$
for oxDNA1, a DNA model with symmetric grooves, was determined in
previous work~\cite{mate15} and found to be in agreement with the ($G=0$)
Moroz-Nelson theory.

For the more realistic oxDNA2, which has the asymmetric grooves of real
DNA and hence twist-bend coupling ($G \neq 0$)~\cite{skor17}, we found
that $C_\text{eff}$ is in excellent agreement with the nonperturbative
theory Eq.~\eqref{eq:ceff_nonp_inv} without any adjustable parameters,
as the elastic constants were determined in previous work~\cite{skor17}.
oxDNA2 appears to be precisely described by our nonperturbative theory
for forces $f \ll f_0$ (recall that oxDNA undergoes internal structural
transitions for forces of a few tens of pN, providing a more stringent
contraint on force than the giant force scale $f_0$).  Put another way,
the ``TWLC plus $G$" is the ``correct" low-force, long-fluctuation
wavelength description of oxDNA2.

\begin{figure}[t]
\centering\includegraphics{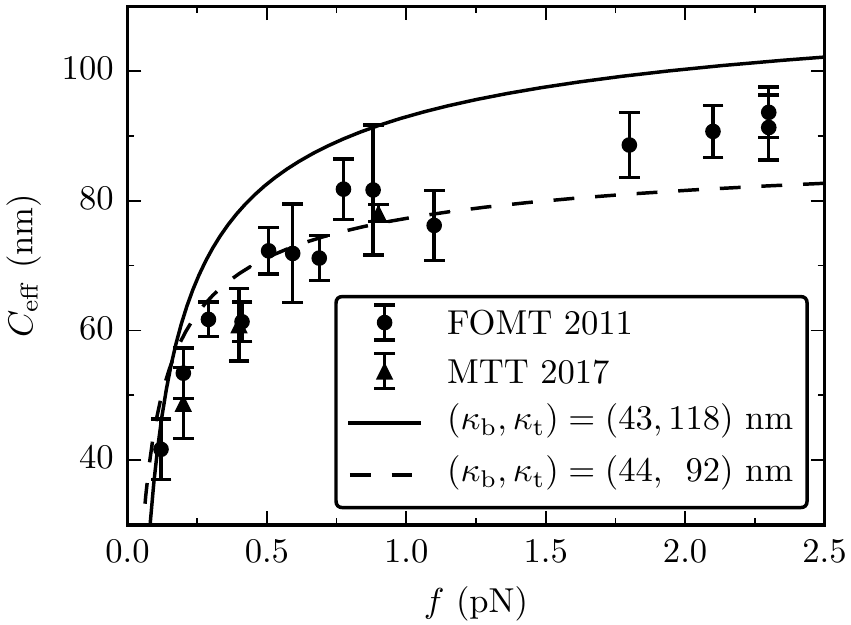}
\caption{Comparison of the theory from Eq.~\eqref{eq:ceff_nonp_inv}
(lines) with Magnetic Tweezer experiments (symbols) for $C_\text{eff}$
vs. force. The lines have the same parametrization of the solid and
dashed lines of Fig.~\ref{Fig:oxdna}, which fit oxDNA and
oxDNA2 data, respectively. Two sets of experiments are shown: the freely-orbiting
magnetic tweezers \cite{lipf11} (circles) and magnetic torque tweezers
\cite{nomi17} (triangles).}
\label{Fig:expt}
\end{figure}

\subsection{Experimental data}
\label{subsec:expt}

We finally compare the analytical results with experimental magnetic
tweezers data of Refs.~\cite{lipf11,nomi17}. Figure~\ref{Fig:expt}
shows experimental data (symbols) together with plots of
Eq.~\eqref{eq:ceff_nonp_inv} for two sets of parameters $\kappa_\text{b}$
and $\kappa_\text{t}$ (lines). The latter are identical to the solid
and dashed lines of Fig.~\ref{Fig:oxdna}, which are numerically precise
descriptions of oxDNA1 and oxDNA2, respectively.  Since the force
fields in oxDNA were carefully tuned to reproduce several mechanical
and thermodynamic properties of DNA \cite{ould10}, it is sensible
to directly compare our nonperturbative theory to experimental data
(Fig.~\ref{Fig:expt}).

As reported in previous papers \cite{lipf10,nomi17} the experimental
$C_\text{eff}$ data are systematically lower than the prediction of
the Moroz and Nelson theory, which precisely matches the oxDNA1 results
(solid line in Fig.~\ref{Fig:expt}). The oxDNA2/nonperturbative theory
(dashed curve) is closer to the experimental data, especially in the
low force regime $f < 1$~pN. However, some systematic deviations are
noticeable at higher forces, where theory appears to underestimate the
experimental $C_{\rm eff}$. In addition, measurements at $f=15$~pN
(albeit for a slightly different assay) yield $C_\text{eff} =
110$~nm~\cite{brya03}, well above the oxDNA2 value of $C_\text{eff} =
92$~nm. 

We conclude by noting that oxDNA2 - which has the realistic features of
groove asymmetry and $G \neq 0$ - produces data in reasonable agreement
with experiments.  We have also shown that in the force range where we
expect that coarse-graining of oxDNA2 should agree with our analytical
results, it does. In that same force range ($f < 10$ pN), oxDNA2 and
our analytical results show some systematic 
deviations from experiments that suggest that physics
beyond simple harmonic elasticity may be in play at intermediate
forces (1 to 10 pN), generating torsional stiffening of DNA.  
A possible mechanism of cooperative structural transition in
a two-state model with different base-pair rise (separation) was
recently discussed in Ref.~\cite{schu15}.
The next generation of coarse-grained DNA models likely will have to
consider this kind of additional, internal degree of freedom to properly
describe severe distortion of DNA by proteins, or similar situations
where strong forces are applied at short length scales.

\subsection{The value of the intrinsic torsional stiffness $C$}

Experiments probing torsional properties of DNA, performed since the 80's, have 
provided estimates of $C$ ranging from $40$~nm to $120$~nm, depending on the 
technique used \cite{shor83,leve86,shib85,fuji90,mosc09,brya12}. All these 
experiments were analyzed within the framework of the standard TWLC, with no 
twist-bend coupling ($G=0$). In the model discussed in this paper with $G \neq 
0$, the bare torsional stiffness $C$, being screened by a large force scale, 
should not be accessible to magnetic tweezers experiments. A main point of this 
paper is that all the experiments ought to measure $\kappa_\text t$ instead. We 
point out that Eq.~\eqref{eq:ceff_nonp} holds for measurements at zero force 
where torsional deformations are governed by the renormalized stiffness 
$\kappa_\text t$ \eqref{kappat} and not by the bare $C$ \cite{nomi17,cara18}. If 
DNA were to follow the model \eqref{mod_any}, torsional measurements at a given 
force should provide a single estimate of the torsional constant $\kappa_\text 
t$, regardless of whether the DNA is under tension or not (provided that $f \ll 
f_0$, which is always the case in experiments).

Deviations of experiments from Eq.~\eqref{eq:ceff_nonp}, observed for $f > 
1$~pN, indicate that DNA is torsionally stiffer for forces in the 1-10 pN 
range than expected from model \eqref{mod_any}. One possibility is that there 
is an additional intrinsic torsional stiffness $C'$ in this regime of forces, 
as postulated by Schurr \cite{schu15}. This explains the large spread in the 
values of the torsional stiffness in earlier experiments: at very weak $f < 
1$~pN or zero forces the torsional behavior is governed by $\kappa_t$ and 
larger forces by a novel torsional constant $C'$. Measurements of torsional 
stiffness from DNA under tension provide systematically higher values compared 
to the zero-tension data (see Table~I of Supplemental of Ref.~\cite{nomi17}), 
suggesting that there are different torsional constants in different force 
ranges. A similar effect was discussed in Ref.~\cite{nomi17}, where it was 
argued that earlier torsional DNA experiments identified two different 
stiffnesses: $\kappa_\text t$ at low/zero tension and $C$, the intrinsic 
stiffness, at high tension. The present paper argues against this conclusion of 
Ref.~\cite{nomi17}, since the intrinsic torsional stiffness is screened at all 
experimentally-accessible forces ($f< f_0$).

\subsection{Effects beyond the TWLC model}
The TWLC with $G\neq 0$ is still a highly-simplified model of DNA molecular 
mechanics.  One might argue that there are degrees of freedom or other features 
of DNA relevant to $C_{\rm eff}$ measurements, which are just not captured by 
the TWLC. The basic TWLC (with $G \neq 0$) Hamiltonian has been obtained at the 
single-base-pair level by coarse-graining detailed molecular MD simulations 
\cite{lank00,lank03}, indicating that the basic symmetry features of the TWLC 
with $G\neq 0$ are present in real DNA, or at least in chemical models of real 
DNA.  In addition to the symmetry properties of base-pair-level deformations, 
the TWLC model also assumes a straight ``zero-temperature'' (non-fluctuating) 
ground state, while real DNA has a sequence-dependent non-straight intrinsic 
shape; evidence for this comes from crystallography of DNA crystals and detailed 
chemical-structural calculations.  Recent work of the latter sort suggests that 
DNA has an appreciable contribution to its effective persistence length by 
sequence-dependent bends \cite{mitc17}.

This leads to the question of whether DNA intrinsic shape 
might contribute appreciably to experimental discrepancies 
between $C_{\rm eff}$ and the predictions of
TWLC-type models. Prior work argues against this, showing that
small-scale random intrinsic bends generate only a simple 
renormalization of the bending modulus (captured in the TWLC model by
shifting the value of $A$) \cite{bens98} and no renormalization of the twisting
modulus \cite{nels98}. 

These theoretical results for small-scale shape disorder nonwithstanding, 
large-scale {\it nonrandom} chiral shape of the molecule (say a coiled shape at 
a length scale $\ell_0$) could give chiral responses at zero temperature 
associated with removal of those coils. 
At finite temperature, effects of such coiling 
would be relevant for forces $< \kB T A/\ell_0^2$, where the correlation 
length for bending fluctuations is large enough to allow fluctuations to be 
affected by $\ell_0$.
Taking $\ell_0 \approx 10$~nm (30~bp) sets this force 
scale to $\lesssim 2$ pN, not far from the force range where experiment and 
TWLC disagree, suggesting that this permanent chiral shape might
contribute to the discrepancy between TWLC and experimental $C_{\rm eff}$ values.  
Future oxDNA-like models, 
which incorporate sequence-shape detail, might be able to observe effects of 
nonrandom chiral structure. From the experimental side, high-precision 
measurements using DNA molecules of different sequence composition (perhaps 
tuned to have nonrandom chiral intrinsic shape) might be able to determine how 
likely it is that sequence is responsible for the discrepancies in $C_{\rm eff}$ 
between experiments and the TWLC theory.

\acknowledgments{Discussions with M.\ Laleman and T.\ Sakaue are
gratefully acknowledged. SN acknowledges financial support from the
Research Funds Flanders (FWO Vlaanderen) grant VITO-FWO 11.59.71.7N,
and ES from KU Leu- ven Grant No.  IDO/12/08. JFM is grateful to the
Francqui Foundation (Belgium) for financial support, and to the US NIH
through Grants R01-GM105847, U54-CA193419 and U54-DK107980.}

\FloatBarrier

\appendix

\section{TWLC at strong stretching}
\label{app:twlc}

We will first consider the simple case of the TWLC ($G=0$), following
closely the approach of Ref.~\cite{mark15}. At high forces, the molecule
is strongly oriented along the force direction, which is chosen to be
parallel to $\unit z$. It proves convenient to decompose the tangent
vector as
\begin{equation}
\unit e_3 = t_z  \unit z + \vec u,
\label{app:defu}
\end{equation}
where the vector $\vec u$ is orthogonal to $\unit z$, i.e.\ $\vec u =
u_x \unit x + u_y \unit y$.  Using the identity $|\unit e_3| = 1 =
\sqrt{t_z^2 + \vec u^2}$ and expanding to lowest order in $\vec u$ we get
\begin{equation}
\unit e_3 = \left[ 1 - \frac{\vec u^2}{2} + \mathcal{O}(\vec u^4) \right] 
\unit z + \vec{u},
\label{app:triv03}
\end{equation}
while its derivative is found to be
\begin{equation}
\frac{\dif\unit e_3}{\dif s} = \frac{\dif\vec u}{\dif s} 
                             + {\cal O}(\vec u^2) \unit z.
\label{app:triv04}
\end{equation}
Combining this with Eq.~\eqref{app:diffeq2}, we find 
\begin{equation}
\Omega_1^2 + \Omega_2^2 = \left( \frac{\dif \unit e_3}{\dif s} \right)^2 
                        = \left( \frac{\dif\vec u}{\dif s} \right)^2 
                        + {\cal O}(\vec u^3).
\end{equation}
Introducing the Fourier transform $\vec{u}_q = \int_0^L \dif s \
e^{-iqs}\, \vec{u} (s)$, and neglecting higher-order terms, we write
the bending and stretching contribution to the energy as follows
\begin{equation}
\frac{A}{2} \!\! \int_0^L \!\!\! \dif s  
\left( \frac{\dif \unit{e}_3}{\dif s} \right)^2\!\!\! - 
\beta \vec{f} \cdot \vec{R} = \frac{1}{2L}
\sum_q \left( A q^2+ \beta f\right) 
\left| \vec{u}_q \right|^2   - \beta f L,
\label{app:bend}
\end{equation}
where we expressed the force as $\vec{f} = f \unit{z}$, while the end-to-end 
vector $\vec R = \int_0^L \dif s \ \unit{e}_3$ was approximated based on
Eq.~\eqref{app:triv03}. 

The torque in Eq.~\eqref{eq:Z0} is coupled to the linking number, which
is the sum of twist and writhe [Eqs.~\eqref{def:twist} and \eqref{def:writhe}, 
respectively]. In the high-force limit, Eq.~\eqref{def:writhe} becomes
\begin{eqnarray}
2 \pi \text{Wr} &=& \frac{1}{2} \int_0^L 
\left(\vec{u} \times  \frac{\dif\vec u}{\dif s}\right) 
\cdot\unit{z} \ \dif s + \mathcal{O}(\vec u^4)
\nonumber \\
&=& \frac{1}{2 L} \sum_q \vec{u}_{q}^{\text T}  
\begin{pmatrix}
0 & - iq \\
iq & 0
\end{pmatrix}
\vec{u}_{q}^{\,*} + \mathcal{O}(\vec u^4),
\label{app_wr}
\end{eqnarray}
where we have rewritten the cross product as a matrix
multiplication. Thus, the writhe couples  the $x$ and $y$ components of
the two-dimensional vector $\vec u_q$.

Adding up all terms, and with the help of simple algebraic manipulations,
we obtain the following energy for the TWLC to lowest order in $\vec u$
\begin{eqnarray}
&&\beta E_0 - \beta \vec{f} \cdot \vec{R} 
- \beta \tau 2\pi \Delta \text{Lk} \nonumber \\
&=& \frac{1}{2L} \sum_q \vec u_q^{\text T} \vec M_q \vec u_q^*
          + \frac{C}{2} \int_0^L \dif s \, \omega_3^2 
          - \frac{\beta^2 \tau^2 L}{2C} - \beta f L, \nonumber \\
\label{app:energy}
\end{eqnarray}
where we introduced the matrix
\begin{equation}
\vec M_q = 
\begin{pmatrix}
 Aq^2+\beta f &-i q \beta \tau \\
 i q \beta \tau  & Aq^2 + \beta f \\
\end{pmatrix}.
\label{app:Mq}
\end{equation}
We have also introduced the shifted twist density 
\begin{equation}
 \omega_3 (s) \equiv \Omega_3 (s) - \frac{\beta \tau}{C},
\label{def:omega3}
\end{equation}
which allowed us to eliminate linear terms in $\Omega_3$. Thus, in the
high-force limit, the TWLC under applied torque reduces, to lowest order,
to a Gaussian model, where bending ($\vec u_q$) and twist ($\omega_3$)
are independent variables. The torque $\tau$ couples to the bending
degrees of freedom through the off-diagonal terms of the matrix $\vec
M_q$. The eigenvalues of $\vec M_q$ are easily found to be
\begin{equation}
\lambda_q^\pm = Aq^2+\beta f \pm q \beta \tau,
\label{defMq}
\end{equation}
and the corresponding eigenvectors are $(\unit x \pm i \unit y)/\sqrt 2$. 
Writing Eq.~\eqref{app:energy} on this basis allows us to calculate the 
partition function, from which the free energy is found to be
\begin{equation}
 F_0 = -f L - \frac{\beta \tau^2 L}{2C} 
     + \frac{\kB T}{2} \sum_q \log \left( \lambda^+_q \lambda^-_q \right),
\end{equation}
where we have neglected additive constants. Expanding to 
quadratic order in $\tau$
\begin{equation}
 \log \left( \lambda^+_q \lambda^-_q \right) \approx 
 \log \left(Aq^2 + \beta f\right)^2 
 - \left(\frac{q\beta \tau}{Aq^2 + \beta f}\right)^2,
\label{another_expansion}
\end{equation}
and replacing the sum over momenta with an integral $\sum_q \to (L/2\pi) 
\int \dif q$, we obtain
\begin{eqnarray}
 F_0(f,\tau) &=& F_0(f,0) - \frac{\beta \tau^2 L}{2C} - 
                 \frac{\beta \tau^2 L}{8A} \sqrt{\frac{\kB T}{f A}}
                 \nonumber\\
             &=& F_0(f,0) - \frac{\beta \tau^2 L}{2C_\text{eff}}.
 \label{eq:F0}
\end{eqnarray}
Combining the two last terms in the right-hand side, one obtains the Moroz and 
Nelson relation [Eq.~\eqref{eq:ceff}]. $F_0(f,\tau=0)$ is the zero-torque free 
energy, and is obtained by integrating the first term at the right-hand side of 
Eq.~\eqref{another_expansion}. Although the integral is divergent, it can be 
regularized by introducing a momentum cutoff $\Lambda \approx 2\pi/a$, where 
$a=0.34$~nm is the separation between neighboring base pairs. As it turns out, 
however, this cutoff does not affect any force-dependent terms, and one has
\begin{equation}
 F_0(f,0) = F_0(0,0) - fL + \sqrt{\frac{f\kB T}{A}} L,
\end{equation}
where $F_0(0,0)$ is cutoff-dependent. Interestingly, from Eq.~\eqref{eq:Z0} one 
finds to lowest order in $\tau$
\begin{equation}
2\pi \Delta \text{Lk} = - \frac{\partial F_0(f,\tau)}{\partial \tau}
= \frac{\beta \tau L}{C_\text{eff}},
\end{equation}
which quantifies the induced over- or undertwisting upon the application 
of a torque. Using this expression, one obtains the force-extension 
relation at fixed linking number~\cite{moro97,moro98}
\begin{equation}
 \frac{z}{L} = -\frac{1}{L}\frac{\partial F_0}{\partial f} 
             = 1 - \sqrt{\frac{\kB T}{4fA}} 
             - \frac{C^2}{2} \left( \frac{\kB T}{4Af}\right)^{3/2}\!
             \left( \frac{2\pi\Delta \text{Lk}}{L} \right)^2,
\end{equation}
which shows a characteristic parabolic profile for the extension of an
over- or undertwisted, stretched molecule.

\section{$\Omega_2$ at strong stretching}
\label{app:omega2}

The thermal average in Eq.~\eqref{eq:pert} contains $\Omega_2$, which
needs to be expressed in terms of $\vec u_q$ and $\omega_3$, the degrees
of freedom of the system [Eq.~\eqref{app:energy}]. For this purpose,
we use the relation
\begin{equation}
 \Omega_2 = \unit e_1 \cdot \frac{\dif\unit e_3}{\dif s},
\label{eq:omega2}
\end{equation}
which can be easily obtained from Eq.~\eqref{app:diffeq2}. In the high-force 
limit, where the tangent $\unit e_3$ points predominantly along the force 
direction, $\unit z$, one has
\begin{equation}
 \unit e_1 = \left[ \cos\psi + {\cal O}(\vec u^2) \right] \unit x + 
             \left[ \sin\psi + {\cal O}(\vec u^2) \right] \unit y +
             {\cal O}(\vec u) \, \unit z.
\label{eq:e1}
\end{equation}
Here we have introduced the twist angle
\begin{equation}
 \psi(s) = \int_0^s \Omega_3(t) \, \dif t + \omega_0 s
         = \int_0^s \omega_3(t) \, \dif t + 
         \left(\omega_0 + \frac{\beta\tau}{C} \right)s,
\label{defpsi}
\end{equation}
and used Eq.~\eqref{def:omega3} to express it in terms of the variable
$\omega_3$. Equation~\eqref{eq:e1} can be obtained by considering
an arbitrary rotation that maps a fixed lab frame triad, e.g.\
$\{ \unit x, \unit y, \unit z \}$, onto the material frame triad
$\{\unit{e}_1, \unit{e}_2, \unit{e}_3\}$ at position $s$, requiring
that $\unit{e}_3$ remains predominantly oriented along the force
direction, as in Eq.~\eqref{app:defu}. Combining Eqs.~\eqref{eq:omega2}
and~\eqref{eq:e1}, it follows that
\begin{equation}
 \Omega_2 
          = \unit\Psi(s) \cdot \frac{\dif \vec u}{\dif s} 
          + {\cal O}(\vec u^3),
\label{app:Omega2}
\end{equation}
where we have defined the unit vector
\begin{equation}
 \unit \Psi(s) \equiv \cos\psi (s)\, \unit x + \sin\psi(s) \, \unit y.
\label{def:Psi}
\end{equation}
Therefore, in the high-force limit, $\Omega_2$ can be written as a scalar
product between a unit vector~$\unit\Psi(s)$, depending exclusively
on twist variables, and a vector~ $\dif\vec u/\dif s$, involving only
the bending degrees of freedom. Finally, from Eq.~\eqref{app:Omega2}
it follows that
\begin{equation}
\widetilde{\Omega}_{2,q} = \frac{1}{L} 
\sum_{q'} (-i q') {\vec \Psi}_{q-q'} \cdot \vec u_{q'}.
\label{FT_Omega2}
\end{equation}
The remainder of the calculation, presented below, will be based upon 
Eqs.~\eqref{app:Omega2} and \eqref{FT_Omega2}. 

\section{Details of the perturbative calculation}
\label{appC}

To calculate the average appearing in Eq.~\eqref{eq:pert}, it first 
needs to be rewritten as a function of the integration variable $\omega_3$ [see 
Eq.~\eqref{def:omega3}]. This can be performed as follows
\begin{eqnarray}
&&\left\langle 
\left(\int_0^L \dif s \, \Omega_2 \Omega_3 \right)^2
\right\rangle_0 = 
\nonumber\\
&=&
\frac{\beta^2 \tau^2}{C^2} 
\left\langle 
\left(\int_0^L \dif s \, \Omega_2 \right)^2
\right\rangle 
+
\left\langle 
\left(\int_0^L \dif s \, \Omega_2 \omega_3 \right)^2
\right\rangle \hspace{0.9cm}
\nonumber\\
&=&
\frac{\beta^2 \tau^2}{C^2} 
\left\langle 
\widetilde{\Omega}_{2,0}^2
\right\rangle +
\frac{1}{L^2} 
\sum_{q,k}
\left\langle 
\widetilde{\Omega}_{2,q}
\widetilde{\Omega}_{2,k}
\widetilde{\omega}_{3,-q}
\widetilde{\omega}_{3,-k}
\right\rangle,
\label{expand}
\end{eqnarray}
where $\widetilde{\Omega}_{2,q}$ and $\widetilde{\omega}_{3,q}$ denote the 
Fourier components of $\Omega_2$ and $\omega_3$, respectively. Note that we have 
neglected a linear term in $\omega_3$, which vanishes due to the symmetry 
$\omega_3 \leftrightarrow -\omega_3$. Moreover, in order to simplify the 
notation, we have dropped the subscript from all averages $\langle.\rangle_0$, 
which will be always calculated within the TWLC model, i.e.\ for $G=0$. 

Before proceeding to the calculation of Eq.~\eqref{expand}, it will prove
useful to first present some properties. In particular, we are going to
use the following expressions, obtained from the correlation functions
in the TWLC model [Eq.~\eqref{app:energy}]
\begin{equation}
 \langle \vec u_q \cdot \vec u_k\rangle = 
 \frac{2L\left( Aq^2+\beta f\right)}
{\left(Aq^2+\beta f\right)^2-\left(q\beta \tau\right)^2}\ \delta_{q,-k}
\label{corru1}
\end{equation}
and
\begin{equation}
\langle \vec{u}_q \otimes \vec{u}_k\rangle = 
\frac{-2iL  q\beta \tau}
{\left(Aq^2+\beta f\right)^2-\left(q\beta \tau\right)^2} \ \delta_{q,-k}.
\label{corru2}
\end{equation}
For convenience, we have introduced the shorthand notation 
\begin{equation}
 \vec a \otimes \vec b \equiv \unit z \cdot \left( \vec a \times \vec b \right)
 = a_x b_y - a_y b_x,
\end{equation}
which is antisymmetric with respect to the interchange of $\vec a$
and $\vec b$. From Eq.~\eqref{corru2} it follows that $\langle \vec
u_q \otimes \vec u_k \rangle =0$, when $\tau=0$ (in this case the
matrix $\vec M_q$ is diagonal, hence the cross-correlations $\langle
u_q^x u_{-q}^y\rangle =0$). Moreover, for $\tau=0$ and $q=-k$,
Eq.~\eqref{corru1} reduces to:
\begin{equation}
\langle \left| \vec u_q \right|^2 \rangle_{\tau=0} = 
\frac{2L}{Aq^2+\beta f},
\label{corrus}
\end{equation}
which can be easily obtained from equipartition~\cite{mark94}.
We are also going to use the following symmetries
\begin{eqnarray}
\langle u_q^x u_k^x\rangle &=& 
\langle u_q^y u_k^y\rangle,
\label{ave_xx} \\
\langle u_q^x u_k^y\rangle &=& -
\langle u_q^y u_k^x\rangle,
\label{ave_xy}
\end{eqnarray}
which allow us to rearrange scalar products as follows
\begin{eqnarray}
&&
\langle \vec\Psi_{-q} \cdot \vec u_q \vec\Psi_{-k} \cdot \vec u_k \rangle = 
\langle u^x_q u^x_k \rangle
\left[
\langle \Psi^x_{-q} \Psi^x_{-k} \rangle +
\right.
\nonumber\\
&&
\left.
\langle \Psi^y_{-q} \Psi^y_{-k} \rangle  
\right]
+ 
\langle u^x_q u^y_k \rangle 
\left[
\langle \Psi^x_{-q} \Psi^y_{-k} \rangle  -
\langle \Psi^y_{-q} \Psi^x_{-k} \rangle
\right]
\nonumber\\
&& =
\frac{1}{2}
\left[
\langle \vec u_q \cdot \vec u_k \rangle
\langle \vec\Psi_{-q} \cdot \vec\Psi_{-k} \rangle
+
\langle \vec u_q \otimes \vec u_k \rangle
\langle \vec\Psi_{-q} \otimes \vec\Psi_{-k} \rangle
\right],
\nonumber\\
\label{c9}
\end{eqnarray}
where we have used the fact that the bending ($\vec u$) and twisting
($\vec\Psi$) degrees of freedom are independent, within the TWLC.  We are
now ready to proceed to the calculation of Eq.~\eqref{expand}. We will
need to evaluate two distinct terms, which will be treated separately.

\subsection{First term in Eq.~\eqref{expand}}

The first term in Eq.~\eqref{expand} already contains a factor of
order $\mathcal{O}(\tau^2)$, which means that up to quadratic order
in $\tau$ it is sufficient to evaluate the corresponding average for
$\tau=0$
\begin{eqnarray}
\left\langle \widetilde{\Omega}_{2,0}^{\,2} 
\right\rangle_{\tau=0} &=&
\frac{1}{L^2} \sum_{q k} (-qk)
\langle 
\vec\Psi_{-q} \cdot \vec u_q 
\vec\Psi_{-k} \cdot \vec u_k 
\rangle_{\tau=0}
\nonumber \\
&=&\frac{1}{2L^2} \sum_q q^2 
\langle | \vec u_q |^2\rangle_{\tau =0}
  \langle 
\left| \vec\Psi_q \right|^2\rangle_{\tau =0},
\label{c6}
\end{eqnarray}
where we have used Eqs.~\eqref{FT_Omega2} and \eqref{c9}, together with the 
property $\langle \vec u_q \otimes \vec u_k  \rangle_{\tau=0} = 0$ [see 
Eq.~\eqref{corru2}]. Next, we need to calculate the following quantity
\begin{equation}
\langle \left| \vec\Psi_q \right|^2 \rangle_{\tau=0} = 
\int_0^L \dif s \dif s' e^{i q (s-s')}
\langle  \unit{\Psi} (s)  \cdot 
\unit{\Psi} (s') \rangle_{\tau=0}.
\label{e1q}
\end{equation}
From Eqs.~\eqref{defpsi} and~\eqref{def:Psi} one finds
\begin{eqnarray}
\langle  \unit{\Psi} (s)  \cdot 
\unit{\Psi} (s') \rangle_{\tau=0} =
\langle \cos [\psi(s)-\psi(s')] \rangle\hspace{1.5cm}
\nonumber\\
=\frac{e^{i\omega_0 (s'-s)}}{2} 
\left\langle 
\exp\left(i \int_s^{s'} \omega_3(t) \, \dif t\right)
\right\rangle + \text{c.c.},\hspace{10pt}
\label{zerom}
\end{eqnarray}
where c.c.\ denotes the complex conjugate. To proceed, we perform a Fourier 
transform of the exponent
\begin{equation}
\int_s^{s'} \omega_3 (t) \,\dif t = \frac{1}{L} 
\sum_q h_q \, \widetilde{\omega}_{3,q},
\label{def_alpha}
\end{equation}
where we have introduced the complex variable
\begin{equation}
h_q = \frac{e^{-iqs'}-e^{-iqs}}{-iq}.
\end{equation}
Performing Gaussian integration in $\widetilde \omega_{3,q}$, one finds
\begin{eqnarray}
\left\langle
\exp 
\left(\pm\frac{i}{L}\sum_q h_q \,
\widetilde\omega_{3,q}\right) 
\right\rangle &=& \exp\left(- \frac{1}{LC} 
\sum_{q>0} h_q h_{-q} \right)
\nonumber\\
&=& \exp\left(-\frac{|s'-s|}{2C}\right).
\label{gauss_shift}
\end{eqnarray}
As expected, the decay of the twist correlation function is governed by
$2C$, i.e.\ the twist persistence length in the TWLC. Combining 
Eqs.~\eqref{zerom},~\eqref{def_alpha} and \eqref{gauss_shift},
we obtain
\begin{equation}
\langle  \unit\Psi(s) \cdot \unit\Psi(s') \rangle_{\tau=0} =
\cos [\omega_0 (s'-s)] \, e^{-|s'-s|/2C}.
\end{equation}
Inserting this in Eq.~\eqref{e1q} yields
\begin{eqnarray}
&&\langle\left| \vec{\Psi}_q \right|^2\rangle_{\tau=0}=
\frac{L}{2} \left(
\frac{1}{i(q-\omega_0)+\frac{1}{2C}} +
\frac{1}{i(q+\omega_0)+\frac{1}{2C}}
\right.\nonumber \\
&&\hspace{2.1cm}+\left. \frac{1}{i(-q+\omega_0)+\frac{1}{2C}} +
\frac{1}{i(-q-\omega_0)+\frac{1}{2C}} 
\right) \nonumber\\
 &&=\frac{L}{2C} 
\left[ 
\frac{1}{(q+\omega_0)^2+\frac{1}{4C^2}} +
\frac{1}{(q-\omega_0)^2+\frac{1}{4C^2}}
\right].
\label{c13}
\end{eqnarray}
Finally, combining Eqs.~\eqref{c6}, \eqref{corrus} and~\eqref{c13} we find
\begin{eqnarray}
\left\langle \widetilde{\Omega}_{2,0}^{\,2} \right\rangle_{\tau=0} &=&
\frac{1}{2 \pi}
\int_{-\infty}^{+\infty} \frac{\dif q \, q^2}{Aq^2 + \beta f}
\left\langle \left| \vec{\Psi}_q \right|^2\right\rangle
\nonumber \\
&=& \frac{L}{A}
\left[1 - \sqrt{\frac{\beta f}{A}} \frac{\sqrt{\frac{\beta f}{A}} + \frac{1}{2C}}
{\left(\sqrt{\frac{\beta f}{A}}+\frac{1}{2C}\right)^2+\omega_0^2}
\right]\nonumber\\
&\equiv& \frac{L}{A} d(f),
\label{zero_mom}
\end{eqnarray}
where we introduced a force-dependent scale factor $d(f)$. Note that $0
\leq d(f) \leq 1$, with $d(f) \to 1$ at small forces and $d(f) \to 0$
at high forces.  Commonly accepted estimates of the DNA elastic constants
put them in the viccinity of $C=100$~nm and $A=50$~nm, while the applied
forces in typical experiments are in the range $0.1~\text{pN} \lesssim
f \lesssim 10$~pN. Recalling that room temperature corresponds to $\kB
T \approx 4$~pN$\cdot$nm, it follows that $\beta f/A$ is at least one
order of magnitude larger than $1/4C^2$. This allows for the following
simplification
\begin{equation}
d(f) \approx \frac{1}{1+f/f_0},
\label{deff}
\end{equation}
neglecting higher-order terms in $C^{-1} \sqrt{A/\beta f}$.  
We have also introduced a characteristic force
\begin{equation}
f_0 = A \kB T\omega_0^2 \approx 600~\text{pN},
\end{equation}
whose value greatly exceeds those at which the double helix
breaks. Thus, for the force range of interest, we may set $d(f)=1$ in
Eq.~\eqref{zero_mom}.  Summarizing, this first term of Eq.~\eqref{expand}
provides the following contribution to the free energy
\begin{equation}
\Delta F^{(1)} = - \frac{G^2}{2} \, \frac{\beta\tau^2}{C^2}
\left\langle \widetilde{\Omega}_{2,0}^{\,2} \right\rangle =
- \frac{\beta \tau^2 L}{2C} \frac{G^2}{AC} d(f).
\label{c17}
\end{equation}
As a final remark, we note that we could have obtained Eq.~\eqref{deff}
using the approximation
\begin{equation}
 \langle \left| \vec\Psi_q \right|^2\rangle 
 \approx \pi L [ \delta(q+\omega_0) + \delta(q-\omega_0) ].
\label{delta_approx}
\end{equation}
Formally, this corresponds to taking the limit $C \to \infty$ in
Eq.~\eqref{c13}, i.e.\ approximating the Lorentzian distributions
$1/[2C(q\pm\omega_0)^2+1/2C]$ with delta functions. This is a valid
approximation as long as $\omega_0 \gg 1/2C$, making the Lorentzians
sharply-peaked at large momenta $q=\pm \omega_0$, where the integrand
in Eq.~\eqref{zero_mom} varies slowly.

\subsection{Second term in Eq.~\eqref{expand}}

Using the same decomposition as in Eq.~\eqref{c9}, the second term in 
Eq.~\eqref{expand} can be written as
\begin{widetext}
\begin{eqnarray}
&&\sum_{q,k} \left\langle 
\widetilde\Omega_{2,q}\, \widetilde\omega_{3,-q}\,
\widetilde\Omega_{2,k}\, \widetilde\omega_{3,-k}
\right\rangle = 
\frac{1}{L^2} \sum_{q,k,q',k'} (-q'k')
\langle
\vec{\Psi}_{q-q'} \cdot \vec{u}_{q'} \vec{\Psi}_{k-k'} \cdot \vec{u}_{k'}
\, \widetilde{\omega}_{3,-q} \widetilde{\omega}_{3,-k}
\rangle
\nonumber \\ 
&&=\frac{1}{2L^2} \sum_{q,k,q',k'} (-q'k')
[
\langle \vec u_{q'} \cdot \vec u_{k'} \rangle
\langle \vec\Psi_{q-q'} \cdot \vec\Psi_{k-k'} \,
\widetilde\omega_{3,-q}
\widetilde\omega_{3,-k}
\rangle
+ \langle \vec u_{q'} \otimes \vec u_{k'} \rangle
\langle \vec\Psi_{q-q'} \otimes \vec\Psi_{k-k'} \,
\widetilde\omega_{3,-q}
\widetilde\omega_{3,-k}
\rangle] 
\nonumber \\
&&=\frac{1}{2L^2} \sum_{q'} q'^2
[\langle |\vec u_{q'}|^2 \rangle
I_\text{s} (q',\tau) +
\langle \vec u_{q'} \otimes \vec u_{-q'} \rangle
I_\text{a} (q',\tau)],
\label{second}
\end{eqnarray}
\end{widetext}
where we used the fact that the $\vec u$ correlators are diagonal in momentum 
space, hence $k'=-q'$ [see Eqs.~\eqref{corru1} and \eqref{corru2}]. We have 
also introduced the symmetric
\begin{equation}
I_\text s (q',\tau) = \sum_{q,k}
\langle 
\vec\Psi_{q-q'} \cdot \vec\Psi_{k+q'} \, 
\widetilde\omega_{3,-q}\,\widetilde\omega_{3,-k}
\rangle,
\label{defIs}
\end{equation}
and antisymmetric products
\begin{equation}
I_\text a (q',\tau) = \sum_{q,k}
\langle \vec\Psi_{q-q'} \otimes \vec\Psi_{k+q'} \,
\widetilde\omega_{3,-q}\,
\widetilde\omega_{3,-k}
\rangle.
\label{defIa}
\end{equation}
In what follows, we are going to compute the contribution of $I_\text s$ 
and $I_\text a$ to the free energy separately.

\subsubsection{Symmetric products}

For the evaluation of Eq.~\eqref{defIs}, we will first focus on the average 
inside the summation, which may be written in the following way
\begin{eqnarray}
&&\langle \vec\Psi_{q-q'} \cdot \vec\Psi_{k+q'} 
\widetilde\omega_{3,-q} 
\widetilde\omega_{3,-k} 
\rangle = 
\nonumber \\
&& 
\int_0^L \dif s \dif s' \, e^{i(q-q')s+i(k+q')s'}
\langle
\unit\Psi(s) \cdot \unit\Psi(s') \, 
\widetilde\omega_{3,-q}
\widetilde\omega_{3,-k} 
\rangle.
\nonumber \\
\label{c23}
\end{eqnarray}
We may now use Eqs.~\eqref{zerom}-\eqref{gauss_shift} so as to obtain
\begin{eqnarray}
&&\langle
\unit\Psi(s) \cdot \unit\Psi(s') \, \widetilde{\omega}_{3,-q}
\widetilde{\omega}_{3,-k} \rangle
\nonumber \\
&&=-\frac{L^2}{2} \left[
e^{i \overline\omega_0 (s'-s)} 
\frac{\partial^2}{\partial h_{-q}\partial h_{-k}}
\left\langle
\exp 
\left(\frac{i}{L}\sum_p h_p \,\widetilde{\omega}_{3,p}\right) 
\right\rangle 
\right.
\nonumber \\
&& \quad 
\left.
+ e^{-i \overline\omega_0 (s'-s)} 
\frac{\partial^2}{\partial h_{-q}\partial h_{-k}}
\left\langle
\exp 
\left(-\frac{i}{L}\sum_p h_p \,\widetilde{\omega}_{3,p}\right) 
\right\rangle
\right] 
\nonumber\\
&& = -L^2 \cos \left[\overline\omega_0 (s'-s) \right] \,
\frac{\partial^2
e^{- \displaystyle \frac{1}{LC} \sum_{p>0} h_{p} h_{-p}}}
{\partial h_{-q}\partial h_{-k}}
\nonumber\\
&& = \frac{L}{C}  \cos \left[\overline\omega_0 (s'-s) \right] \,
\left( \delta_{q,-k} - \frac{1}{LC} h_q h_k
\right)
e^{-|s'-s|/2C},
\nonumber\\
\label{c24}
\end{eqnarray}
where we have introduced the shifted intrinsic twist
\begin{equation}
\overline\omega_0 \equiv \omega_0 + \frac{\beta\tau}{C}.
\label{def:barom}
\end{equation}
Differently from the calculation of $\langle \Omega_{2,0}^2 \rangle$ in 
Eq.~\eqref{c6}, we can no longer ignore the torque dependence of $\psi$ 
[see Eq.~\eqref{defpsi}]. Plugging Eq.~\eqref{c24} back in Eq.~\eqref{c23}, 
integrating in $s$ and $s'$ and summing over $q$ and $k$, we obtain
\begin{eqnarray}
&&I_\text{s}(q',\tau) 
= \frac{L^3}{C} - \frac{L^3}{4C^2} \int_{-\infty}^{+\infty} \dif r
\cos(q'r) \cos(\overline\omega_0 r)  \, e^{-|r|/2C} 
\nonumber\\
&&
= \frac{L^3}{C} - \frac{L^3}{8C^3} 
\left[
\frac{1}{(q'+\overline\omega_0)^2 + \frac{1}{4C^2}} +
\frac{1}{(q'-\overline\omega_0)^2 + \frac{1}{4C^2}}
\right]
\nonumber\\
&&\approx \frac{L^3}{C} - \frac{\pi L^3}{4C^2}
\left[ \delta\left( q+\overline\omega_0 \right)
+ \delta\left( q-\overline\omega_0 \right) \right].
\label{psi2omega2}
\end{eqnarray}
Throughout the calculation we introduced the variable $r \equiv s'-s$ in the 
double integral. Similar to Eq.~\eqref{delta_approx}, we also approximated the 
two Lorentzians with delta functions. Note that $I_\text s$ depends on the 
torque $\tau$ through $\overline\omega_0$, as indicated by 
Eq.~\eqref{def:barom}. Combining Eqs.~\eqref{corru1} and~\eqref{psi2omega2}, one 
finds
\begin{eqnarray}
&&\frac{1}{2L^2} \sum_{q} q^2 
\left\langle \vec{u}_{q} \cdot \vec{u}_{-q} \right\rangle
I_\text{s}(q,\tau)
\nonumber\\
&&
= \frac{1}{L} \sum_q
\frac{Aq^2+\beta f}
{\left(Aq^2+\beta f\right)^2-\left(q\beta \tau\right)^2}
\, q^2 I_\text{s}(q,\tau)
\nonumber\\
&&
= \frac{1}{L} \sum_q \frac{q^2 I_\text{s}(q,\tau)}{Aq^2+\beta f}
+ \frac{\beta^2 \tau^2}{L} \sum_q 
\frac{q^4 I_\text{s}(q,0)}{(Aq^2+\beta f)^3}
+ {\cal O}(\tau^4).
\nonumber\\
\label{c27}
\end{eqnarray}
We are interested in terms proportional to $\tau^2$. There are two
such contributions, the first one being 
\begin{eqnarray}
&&\frac{1}{L} \sum_q \frac{q^2 I_\text s(q,\tau)}{Aq^2+\beta f} = 
\frac{-L^3}{AC} \left( \sqrt{\frac{\beta f}{4A}}+ \frac{1}{4C} 
\frac{\overline\omega_0^2}{A\overline\omega_0^2 + \beta f}+\ldots\right)
\nonumber\\
&& = - \frac{L^3}{AC} \sqrt{\frac{\beta f}{4A}}
-\frac{L^3 d(f)}{4 AC^2} \times
\nonumber\\
&&\quad\times\left\{
1 + \frac{\beta \tau}{C\omega_0} 
\widetilde{d}(f) 
+ \frac{\beta^2 \tau^2}{C^2\omega_0^2} 
\widetilde{d}(f)  [4d(f)-1]
\right\} + \ldots \nonumber\\
&&= - \frac{L^3}{AC} \sqrt{\frac{\beta f}{4A}}
+ \ldots,
\label{c29}
\end{eqnarray}
where we defined $\widetilde d(f) \equiv 1 - d(f)$, with $d(f)$ the
scale factor given in Eq.~\eqref{deff}, and where the dots indicate
omitted terms, which do not significantly contribute to the result. These
terms are either independent of the torque and force, or are of higher
order than $\tau^2$.  We note that $\widetilde{d}(f) \ll 1$, i.e.\
it is negligibly small for the experimentally-accessible forces $f
\ll f_0 \approx 600$~pN. The only surviving term in Eq.~\eqref{c29} is
independent of $\tau$ and proportional to $\sqrt{f}$, hence contributing
to the force-extension response. The remaining term to evaluate in
Eq.~\eqref{c27} is
\begin{eqnarray}
&&
\frac{\beta^2 \tau^2}{L}  
\sum_q \frac{q^4 I_\text{s}(q,0)}{(Aq^2+\beta f)^3}
= \frac{\beta^2 \tau^2}{4\pi} 
\frac{\partial^2}{\partial A^2}
\int \frac{\dif q}{Aq^2+\beta f}I_\text{s}(q,0)
\nonumber\\
&&
= \frac{\beta^2 \tau^2 L^3}{C} \,
\left[ \frac{3}{16 A^2} \sqrt{\frac{\kB T}{fA}} - 
\frac{2 d^3(f)}{CA^3 \omega_0^2} \right] 
\nonumber\\
&&\approx \frac{3 \beta^2 \tau^2 L^3}{16 A^2 C} \,
\sqrt{\frac{\kB T}{fA}}.
\label{c30}
\end{eqnarray}
Note that terms containing $\omega_0^{-2}$ are always multiplied by $A^{-2}$ or 
$C^{-2}$, hence forming dimensionless constants. Typical values for the case 
of DNA are $(A\omega_0)^{-2} \approx 10^{-4}$ and $(C\omega_0)^{-2} \approx 3 
\times 10^{-5}$, which provide negligible contributions to the free energy, 
compared to other terms of the same order in $\tau$. Therefore, the term 
proportional to $d^3$ in Eq.~\eqref{c30} can be safely neglected. 
Combining Eqs.~\eqref{c27}-\eqref{c30}, we find that the relevant contribution 
of the symmetric term in Eq.~\eqref{second} to the free energy is 
\begin{equation}
\Delta F^{(2)} = \frac{G^2 L}{4AC} \sqrt\frac{f\kB T}{A}
- \frac{\beta \tau^2 L}{2C} \frac{3G^2}{16A^2} 
\sqrt{\frac{\kB T}{fA}}.
\label{DeltaF2}
\end{equation}

\subsubsection{Antisymmetric products}

The final part of the derivation is devoted to the calculation of the 
antisymmetric product in Eq.~\eqref{second}. We start by expanding 
Eq.~\eqref{corru2} as follows
\begin{equation}
\langle \vec u_q \otimes \vec u_{-q} \rangle =
\frac{-2iL  q\beta \tau } {\left(Aq^2+\beta f\right)^2} + \mathcal O (\tau^3).
\label{ucor_app}
\end{equation}
The calculation of the twist correlator is performed in a similar fashion as 
above, which yields
\begin{eqnarray}
&&\langle
\unit\Psi(s) \otimes \unit\Psi(s') \, 
\widetilde{\omega}_{3,-q}
\widetilde{\omega}_{3,-k} \rangle = 
\nonumber \\
&& \frac{L}{C}  \sin \left[\overline\omega_0 (s'-s) \right] \,
\left( \delta_{q,-k} - \frac{1}{LC} h_q h_k
\right)
e^{-|s'-s|/2C}.
\nonumber\\
\end{eqnarray}
We may take the Fourier transform of this expression, and plug it back 
into Eq.~\eqref{defIa}, so as to obtain
\begin{eqnarray}
&&I_\text{a}(q',\tau) 
= -\frac{iL^3}{8C^3} 
\left[
\frac{1}{(q'+\overline\omega_0)^2 + \frac{1}{4C^2}} -
\frac{1}{(q'-\overline\omega_0)^2 + \frac{1}{4C^2}}
\right]
\nonumber\\
&&\approx -\frac{i\pi L^3}{4C^2} 
\left[ \delta\left( q'+\overline\omega_0 \right)
- \delta\left( q'-\overline\omega_0 \right) \right].
\label{psi2omega3}
\end{eqnarray}
Finally, plugging Eqs.~\eqref{ucor_app} and \eqref{psi2omega3} into the
second term of Eq.~\eqref{second}, transforming the sum into an integral
and performing the remaining integration, we find
\begin{eqnarray}
&&\frac{1}{2L^2} \sum_{q} q^2
\left\langle \vec{u}_{q} \otimes \vec{u}_{-q} \right\rangle
I_\text{a}(q,\tau)
\nonumber\\
&&
\approx-\frac{i}{L} \sum_q \frac{q\beta \tau}
{\left(Aq^2+\beta f\right)^2} \, q^2 I_\text{a}(q,\tau)
 \nonumber\\
 &&
= \frac{i \beta \tau}{2\pi} \frac{\partial}{\partial A} \int 
\frac{\dif q q I_\text{a}(q,\tau)}{Aq^2+\beta f} 
= \frac{\beta \tau L^3}{4C^2} \frac{\partial}{\partial A}
\frac{\overline \omega_0}{A \overline \omega_0^2 + \beta f}
\nonumber\\
&& 
= -\frac{\beta \tau L^3}{4C^2}
\frac{\overline \omega_0^3}{\left(A \overline \omega_0^2 + \beta f\right)^2}
\approx
- \frac{\beta \tau d^2(f) L^3}{4A^2C^2\omega_0},
\label{c31}
\end{eqnarray}
where we have omitted terms, which are either higher order in $\tau$, or 
negligibly small compared to other terms of the same order [recall 
$(C\omega_0)^{-2} \approx 3\times 10^{-5}$].
Summarizing, the contribution of the antisymmetric product to the free energy is
\begin{equation}
\Delta F^{(3)} = \frac{G^2 d^2(f) \tau L}{8 A^2 C^2 \omega_0}.
\label{c36}
\end{equation}

\subsection{Collecting the results}

Throughout the derivation we found three distinct contributions to
the free energy, coming from Eqs.~\eqref{c17}, \eqref{DeltaF2} and
\eqref{c36}.  Adding these to Eq.~\eqref{eq:F0}, i.e.\ the free energy
of the TWLC, we find
\begin{equation}
\frac{1}{L} F(f,\tau) \approx -f + \sqrt{\frac{f \kB T}{A_\text{eff}}} 
+ \Gamma \tau
- \frac{\beta \tau^2}{2C_\text{eff}},
\label{app:Fpert}
\end{equation}
where we have omitted both terms independent of $f$ and $\tau$ and 
higher-order corrections in $\tau$ and $G$. We have also introduced the 
effective bending stiffness
\begin{equation}
\frac{1}{A_\text{eff}} =
\frac{1}{A} \sqrt{1 + \frac{G^2}{4AC}} \approx 
\frac{1}{A} \left(1+ \frac{G^2}{2AC} \right),
\label{aeff_exp}
\end{equation}
together with the proportionality constant
\begin{equation}
\Gamma = \frac{G^2 d^2(f)}{8A^2C^2 \omega_0}.
\label{app:defGamma}
\end{equation}
Finally, we reach the following expression for the effective torsional 
stiffness
\begin{equation}
\frac{1}{C_\text{eff}} = 
\frac{1}{C} \left[1 + \frac{G^2}{AC} \, d(f) \right]+ 
\frac{1}{4A} \left(  1+ \frac{3G^2}{4AC} \right) 
\sqrt{\frac{\kB T}{f A}},
\label{ceff_exp}
\end{equation}
corresponding to Eq.~\eqref{eq:ceff_pert} of the main text, and the 
central result of this work.

\section{Intrinsic bending}\label{sec:int_bend}

The analysis above is based on description of the
ground-state configuration of DNA relative to a straight 
molecular axis
[$\vec\Omega = \vec 0$ in Eq.~\eqref{app:diffeq2}], i.e., for 
a molecular axis which is straight in the ground state.
However, one can also choose coordinates where the ground
state of the double helix is a helix while still respecting the
symmetry of the elastic model.  In fact, this is a rather
natural outcome for most choices of DNA deformation which
are based on molecular modeling, where coordinates are usually
chosen relative to the orientations of the base pairs 
(e.g., using the vector connecting the junctions of the bases
to the sugar-phosphate backbone as a reference), due to
the groove asymmetry of DNA.  Most relevant here,
our previous determination of the elastic constants
of oxDNA2 \cite{skor17} analyzed deformations relative to
a helical coordinate system.  We now show how to transform
the elastic constants in such a helical coordinate system
to the straight-line coordinates relevant to our calculations.
 
Intrinsic bending consistent with groove asymmetry, 
usually reported in the DNA literature as a nonzero value of
the average roll~\cite{pere08},
can be described using the following
modification of Eq.~\eqref{app:diffeq2}
\begin{equation}\label{eq:rot_intr_bend}
 \frac{\dif\unit e_i}{\dif s} = 
(\vec\Omega + l_2\unit e_2 + l_3\unit e_3) 
\times \unit e_i,
\end{equation}
where $l_2$ and $l_3$ correspond to the intrinsic bending and
twisting densities, respectively, with $l_2 \ll l_3$. 
A nonzero $l_1$ is incompatible with the symmetry of the double helix.

Solving Eq.~\eqref{eq:rot_intr_bend} for $\vec \Omega = \vec 0$, one finds that
the ground-state configuration is a helix, with a linking number
equal to $\text{Lk}_0 = \omega_0 L/2\pi$, where
\begin{equation}\label{eq:om0_l2_l3}
 \omega_0 = \sqrt{l_2^2+l_3^2}.
\end{equation}
Furthermore, from the solution of Eq.~\eqref{eq:rot_intr_bend}, it
follows that the rotation matrix transforming the helical ground
state of Eq.~\eqref{eq:rot_intr_bend} to the straight one of
Eq.~\eqref{app:diffeq2} is
\begin{equation}\label{eq:rot_mat}
 \vec R = 
 \begin{pmatrix}
  1 & 0 & 0 \\
  0 & l_3/\omega_0 & l_2/\omega_0 \\
  0 & -l_2/\omega_0 & l_3/\omega_0
 \end{pmatrix},
\end{equation}
expressed on the body frame $\{ \unit e_1, \unit e_2, \unit e_3 \}$
of the former.

The total elastic free energy should not depend on the coordinate system
used to describe it, so the energy in the helical coordinates ($l_2,
l_3 \neq 0$) should equal that found in non-helical coordinates ($l_2 =
0$ and $l_3 = \omega_0$) The deformations in the ``straight" model are
given by $\vec \Omega' = \vec R^\text T \vec \Omega$, where $\vec\Omega$
are the deformation parameters of the helical model. From the condition
that the integrand of Eq.~\eqref{mod_any} has to remain invariant under
this transformation, one obtains the following relations mapping the
elastic constants from the helical coordinates to the straight ones:
\begin{align}
 A_{1,{\rm s}} &= A_1, \label{eq:A1_rot}\\[5pt]
 A_{2,{\rm s}} &= A_2 - \frac{2xG - x^2(C-A_2)}{1+x^2}, \label{eq:A2_rot}\\
 C_{\rm s}     &= C   + \frac{2xG - x^2(C-A_2)}{1+x^2}, \label{eq:C_rot}\\
 G_{\rm s}     &= G - \frac{x(C-A_2) + 2x^2G}{1+x^2}, \label{eq:G_rot}
\end{align}
where $x \equiv l_2/l_3 \ll 1$, and where the subscript s indicates the
``straight" frame result. The transformation (\ref{eq:rot_mat}) mixes
$A_2$, $G$ and $C$, changing their values, but conserves the symmetry
of the elastic constant matrix.  These formulae allow one to measure
elastic constants using arbitrarily chosen helical reference coordinates,
and then convert them to elastic constants suitable for using strains
defined relative to a straight-line ground state.

For unconstrained (zero force and torque) oxDNA2 simulations, we
measured reference helix parameters $l_2 = 0.1349$ nm$^{-1}$ and $l_3 =
1.774$ nm$^{-1}$, giving $\omega_0 = 1.779$ nm$^{-1}$ and $x =  0.076$.
Elastic constants reported in Ref. \cite{skor17} ($A_1 = 85$ nm, $A_2 =
39$ nm, $C = 105$ nm, and $G = 30$ nm) were measured in reference to
helical coordinates; for use in our analytical theory we transform them
to the straight coordinates using (\ref{eq:A1_rot})-(\ref{eq:G_rot})
to obtain the oxDNA2 values in Table 1.


%
\end{document}